\documentclass[nolinenumbers, twocolumn]{aastex631}

\usepackage{eqnarray, amsmath}
\usepackage{mathtools}

\begin{document}

\title{Role of magnetic arcades in explaining the puzzle of the gamma-ray emission from the solar disk}

\newcommand{\ff}{\textcolor{red}}
\newcommand{\ep}{\textcolor{magenta}}

\author[0009-0009-5314-348X]{Eleonora Puzzoni}
\affiliation{Lunar and Planetary Laboratory, University of Arizona, 1629 E University Blvd, Tucson, AZ 85721, USA}

\author{Federico Fraschetti}
\affiliation{Lunar and Planetary Laboratory, University of Arizona, 1629 E University Blvd, Tucson, AZ 85721, USA}
\affiliation{Center for Astrophysics, Harvard \& Smithsonian, 60 Garden Street, Cambridge, MA 02138, USA}

\author{József Kóta}
\affiliation{Lunar and Planetary Laboratory, University of Arizona, 1629 E University Blvd, Tucson, AZ 85721, USA}

\author{Joe Giacalone}
\affiliation{Lunar and Planetary Laboratory, University of Arizona, 1629 E University Blvd, Tucson, AZ 85721, USA}



\begin{abstract}

The interpretation of gamma-ray emission originating from the solar disk ($0.5^\circ$ in angular size) as due to the interaction of Galactic Cosmic Rays (GCRs) with the solar atmosphere has remained a central challenge in solar physics. After the seminal work by Seckel, Stanev, and Gaisser (SSG91) based on GCRs magnetic mirroring, discrepancies between models and observations persist, indicating the need for a novel approach. The present work focuses on exploring the impact of a closed magnetic field geometry in the low photosphere on the observed gamma-ray flux. We track numerically with the PLUTO code the trajectories of test-particle protons within a static $\sim 20$ Mm scale height magnetic arcade adjacent to jets. By making use of numerical vertical density profiles we inject particles at distinct chromospheric/photospheric altitudes, mimicking the migration of GCRs from neighboring flux tubes into closed arcades. Remarkably, our model reproduces a flat gamma-ray spectrum below $\sim 33$ GeV, a nearly-isotropic emission at $\sim 10$ GeV, both consistent with Fermi-LAT observations, and a near-limb emission at $\sim 1$ TeV.
Our model can also reproduce the flux-drop detected by HAWC ($\sim 1$ TeV). Finally, we argue that the spectral dip observed at $\sim$ 40 GeV may result from the flux suppression at low energy due to the cross-field diffusion, which would produce a cutoff. These findings underscore the pivotal role of closed magnetic field structures in shaping the solar disk gamma-ray emission. 

\end{abstract}



\section{Introduction} 
\label{sec:intro}

\subsection{Sun as a quiescent source of gamma rays}
The Sun is a brilliant quiescent source of gamma-rays, shining from both its halo and disk \citep{Orlando2008}: $\sim 20^\circ$ \citep{Abdo2011} and $\sim 0.5^\circ$ \citep{Linden2018} in angular size, respectively. 
In the latter region, gamma-ray emission might arise from the decay of neutral pions produced by multi-GeV or TeV GCRs hadrons interacting with protons within the solar atmosphere.
\cite{Dolan1965} were pioneers in suggesting that the gamma-ray emission from the quiescent Sun could stem from the interaction between GCR protons and protons within the solar atmosphere. Moreover, the gamma-ray emission from the halo originates from the Inverse Compton (IC) scattering  of the GCRs electrons and positrons upon the optical photon field of the near-Sun environment \citep[see, e.g.,][]{Moskalenko2006, Orlando2007, Orlando2021}. Additionally, energetic transient solar events, such as solar flares, have the potential to accelerate particles, yielding  $\lesssim$ GeV gamma rays via hadronic collisions \citep[][]{Kafexhiu.etal:18,Tang2018}.

It is noteworthy that, apart from the Sun, steady gamma-ray emission has not been detected from other individual stars.
However, an analysis conducted by \cite{deMenezes2021} established an upper limit of $3.3 \times 10^{-11}$ ph cm$^{-2}$ s$^{-1}$  for energies $>500$ MeV on stellar Fermi/LAT gamma-ray fluxes, consequently constraining the average electron local density in the vicinity of the nine superluminous stars within $600$ pc from the Sun to be less than twice that observed in the Solar System.

\subsection{Early modeling}
The first theoretical model for the high-energy emission from the solar disk was proposed by \cite{Seckel1991} (hereafter SSG91), wherein the pivotal role of the solar magnetic field was underscored. 
\cite{Tang2018} expounded that unabsorbed gamma rays could solely emanate from cosmic rays skimming the solar surface and directed toward Earth when magnetic fields are not taken into account, resulting in diminished gamma-ray emission from the central region of the solar disk \citep[see][]{Zhou2017}.
Consequently, the magnetic field regulates the spatial distribution of the gamma ray emission from the disk \citep[][]{Albert2023}. 
SSG91 concept for augmenting the gamma-ray flux generated by the solar disk involves mirroring of multi-GeV to TeV GCR protons deeper into the solar atmosphere, and confined within magnetic flux tubes.
\cite{Linden2022} point out that cosmic ray protons must be deflected at the appropriate depth of the solar atmosphere to encounter sufficient matter for gamma-rays production while avoiding absorption by the Sun.

SSG91 posited that with advancements in technology, the gamma-ray flux predicted by their model could be detected.
\cite{Hudson1989} was the pioneer in suggesting that the sensitivity of the Energetic Gamma Ray Experiment Telescope (EGRET), onboard the Compton Gamma Ray Observatory spanning the energy range from $\sim 30$ MeV to over $20$ GeV \citep{Gehrels1993}, would enable the observation of gamma-ray emission from quiet Sun, beside flaring activity.
\cite{Thompson1997} analyzed the EGRET data, demonstrating that the Moon, observed by the telescope multiple times between 1991 and 1994, serves as a high-energy gamma-ray source with flux values varying in accordance with the solar cycle.
Nevertheless, they contend that EGRET was unable to detect the quiet Sun.

\subsection{The observational puzzles}
\label{sec:puzzles}
Despite the complexity due to the presence of other gamma-ray sources in the galaxy, \cite{Orlando2008} conducted a thorough analysis of EGRET data marking the first detection of the quiet Sun in gamma-rays. Their study discerned two distinct emission components: a hadronic component originating from the solar disk and leptonic component resulting from IC scattering of CR electrons off the solar radiation field. 
Subsequent to its launch in 2008, the NASA Fermi Gamma-Ray Space Telescope, equipped with the Large Area Telescope (LAT), an imaging gamma-ray detector operating in the range of $20$ MeV to $300$ GeV, brought a paradigm shift in observations. 
\cite{Atwood2009} and \cite{Giglietto2012} meticulously examined the two emission components identified by Fermi-LAT, attaining high statistical significance.
\cite{Abdo2011} reported an integral flux (within the range of $0.1$ to $10$ GeV) of the extended IC emission up to a radius of approximately $20^{\circ}$ around the Sun, in reasonable agreement with predictions by \cite{Orlando2008}. 

On the contrary, the predictions of the SSG91 model began facing scrutiny. 
In particular, these are the observational points in tension with SSG91 predictions and the new observed and unexpected features:

\begin{itemize}
\item \textbf{Observed gamma-ray spectrum extends beyond SSG91 cutoff}. 
Notably, the SSG91 model predicts an abrupt cutoff at $E \simeq 5$ GeV, while the observed gamma-ray flux from the solar disk extends up to $100$ GeV \citep{Linden2018} and beyond \citep[$\gtrsim 200$ GeV, in][]{Tang2018}.
This was later confirmed by \cite{Albert2023}, which observed a gamma-ray flux up to $\sim 2.6$ TeV with HAWC.
The predicted cutoff in the SSG91 model might be attributed to the absence of confinement mechanisms, such as flux tubes or other assumed trapping structures, for higher-energy GCRs.

\item \textbf{Disk flux excess relative to SSG91 predictions}.
During the initial 18 months of Fermi-LAT operations (Fermi Pass 6 observations), \cite{Abdo2011} reported an observed integral flux (above $100$ MeV) from the solar disk approximately seven times higher than predicted by the SSG91 model, particularly during the minimum of solar cycle $24$ (from August 2008 to late 2019).
The discrepancy is likely rooted in assumptions about the structure of the interplanetary magnetic field or the assumed opacity for the hadronic interaction of GCR with atmospheric protons (up to the case that every GCR produces gamma-rays). 
TeV observations of the Sun's shadow by the Tibet Array, spanning the years 2000 to 2009, have been leveraged to infer the structure of the near-Sun interplanetary magnetic field \citep[see][]{Amenomori2018}, where high-energy cosmic rays are obstructed and deflected from radial direction, giving rise to the Sun shadow \citep[][]{Becker2020}.
The extended mission duration of Fermi (2008 to 2020, covering two solar minima) allowed \cite{Linden2022} to confirm this finding.

\item \textbf{Anti-correlation between the solar activity and the GeV solar gamma-ray flux}. 
\cite{Ng2016} first brought to light that the gamma-ray flux tends to be greater during solar minimum compared to solar maximum. 
Subsequent research by \cite{Linden2018}, based on Pass 8 data, further corroborated this anti-correlation up to $\sim 30$ GeV.
This noteworthy discovery was subsequently validated through a comprehensive analysis of the entire solar cycle by \cite{Linden2022}.
Their results underscored that, during solar minimum, the gamma-ray flux can be up to a factor of at most 2 larger compared to solar maximum. 
In contrast, a much larger oscillation in the GCRs flux is observed at 1 AU during the solar cycle, associated with the change of sunspot number.

\item \textbf{Spatial distribution of the gamma-ray emission from solar disk}.
In the study conducted by \cite{Linden2022}, it was established that the spatial distribution of emitted gamma-rays within the solar disk is contingent upon their energy. Specifically, lower energy emissions (ranging from $10$ to $50$ GeV) is isotropic, whereas higher energy emissions (exceeding $50$ GeV) is less isotropic and with a time-dependent anisotropy.
Recently, \cite{Arsioli2024} conducted an analysis of Fermi-LAT data spanning from August 2008 to January 2022. 
The findings revealed an asymmetric gamma-ray emission pattern emanating from the solar disk during the peak of solar cycle 24. 
Notably, during the solar maximum, gamma-ray emissions at higher energies (ranging from $20$ to $150$ GeV) are predominantly observed at the solar south pole region, while emissions at lower energies (ranging from $5$ to $20$ GeV) are predominantly observed at the solar north pole region.
A possible explanation for the stronger emission from the limb at energies above $\sim$ $1$ TeV was previously given by \cite{Zhou2017}: high-energy GCRs graze the Sun producing escaping gamma-rays.

\item \textbf{Comparing Hardness in Gamma-Ray and Cosmic Ray Spectra}.
Finally, the observed gamma-ray ($\sim 1 -10^3$ GeV) spectrum is significantly harder than the impinging GCRs proton spectrum, as elaborated in prior studies \citep[see also][]{Orlando2008, Mazziotta2020}. 
In particular, the gamma-ray emission detected from the solar disk above $1$ GeV, as reported by \cite{Mazziotta2020}, exhibits a slope of approximately $-0.2$, in contrast with the steeper slope of the GCR spectrum $\sim -0.7$. 
In the SSG91 model, the slope value falls within the range defined by the slopes of the observed gamma-ray flux and the cosmic ray spectrum.

\item \textbf{Dip in the gamma-ray spectrum}.
\cite{Linden2018}, \cite{Tang2018}, and \cite{Linden2022} discovered new features that were not anticipated, including a statistically significant dip in the energy range of $30-50$ GeV \citep[see, e.g.,][]{Tang2018} that persists both during and after the solar minimum.
The CALorimetric Electron Telescope (CALET), launched in August 2015 (and operating from the International Space Station Japanese Experiment Module Exposed Facility), observes from $1$ GeV up to $10$ TeV and did not detect this spectral dip. 
Nevertheless, the statistical errors do not allow to rule it out \citep[][]{Cannady2022}.
Models presented by \cite{Gutierrez2022}, incorporating the energy-dependent shadow of the Sun measured by the High-Altitude Water Cherenkov
Gamma-ray Observatory (HAWC), and \cite{Li2024}, considering a flux tube and a flux sheet (lacking closed magnetic field lines) for reflecting cosmic rays, could not reproduce the dip. However, \cite{Li2024} determined that a two-zone model might produce a flux decrease at energies  $\sim 40$ GeV.

\end{itemize}

\subsection{Developing a new theoretical model}
The Sun and the heliosphere navigate through interstellar space, traversing a nearly isotropic sea of GCRs. 
GCRs with energies reaching up to $\gtrsim 10^{15}$ eV are thought to originate from shocks in supernova remnants \citep[see, e.g.,][]{Lee2000}. 
The local interstellar spectrum of GCR ions below $100$ GeV$/$nucl closely adheres to a single power-law, exhibiting a roll-over at lower energy, i.e., $\sim 100$ MeV$/$nucl, \citep[][]{Cummings2016} due to ionization losses \citep[][]{Ip1985}.
In the explanation of the Fermi/LAT observations of the solar disk \citep{Linden2018}, the configuration of the solar magnetic field plays an important role in the transport and confinement of GCRs. 
Despite efforts such as the SSG91 flux tube model, existing models are far from elucidating specific characteristics observed in the gamma-ray flux.
\cite{Gutierrez2022} reproduced a gamma-ray spectrum flattening at $E_\gamma < 100$ GeV solely relying on HAWC data without incorporating any specific model for the magnetic field geometry in the solar atmosphere. \cite{Li2024} focused on two-zone model with flux tubes (i.e., open magnetic field lines) and flux sheets.
Notably, none of these model incorporated closed magnetic field lines, which may play a crucial role in the trapping of GCRs and subsequently influence gamma-ray production. 
Indeed, \cite{Arsioli2024} assert a connection between the magnetic field, gamma-ray emission anisotropies, and the solar cycle.
Consequently, there arises a compelling need for a novel theoretical framework, featuring an alternative expression for the magnetic field.

\begin{figure}
    \centering
    \includegraphics[width=0.5\textwidth]{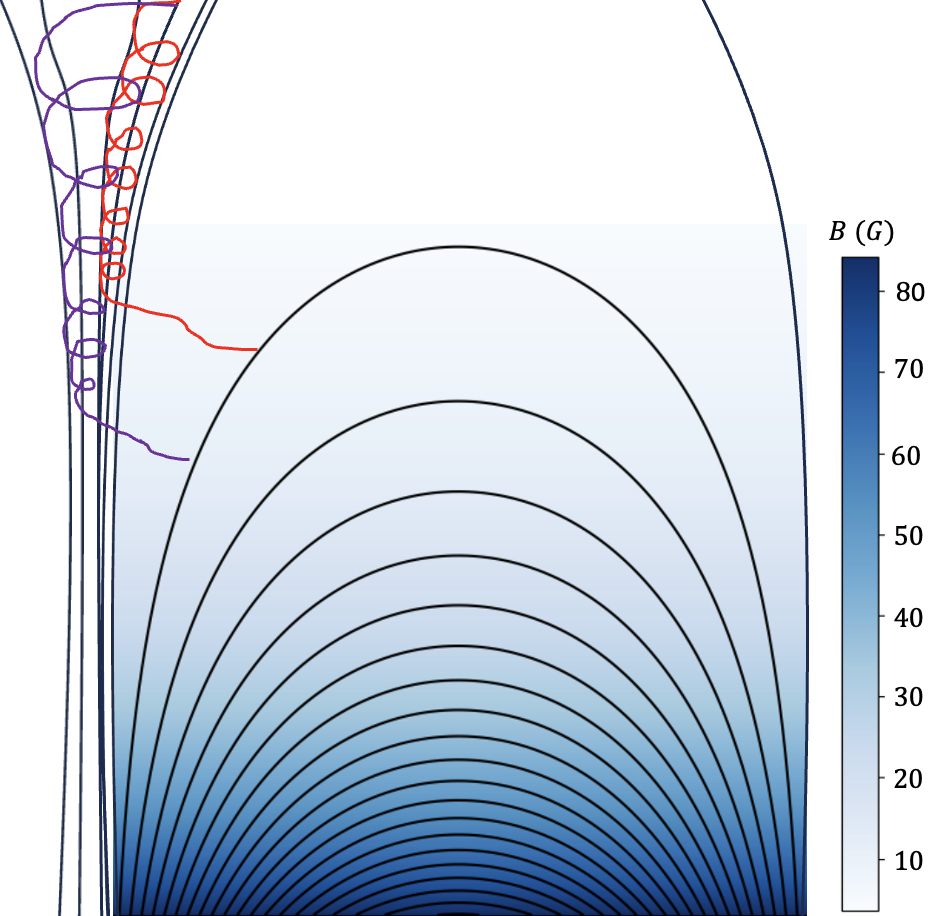}
    \caption{Cartoon of the trajectory of an approaching relatively low and high energy GCR (in red and purple, respectively) gyrating first along the open field lines before entering the closed magnetic arcade. The magnetic field strength is blue-white color-scaled. Solid black lines represent the magnetic field lines $A_z$ (see Equation \ref{eq:Az}).}. 
    \label{fig:open_field}
\end{figure}

In this work we propose a comprehensive investigation of the role of closed magnetic field structures in the entrapment of GCRs within the dense solar atmosphere after they access the magnetic arcades via the cross-field diffusion
\citep[discussed in, e.g.,][]{Chuvilgin1993, Bieber1997, Giacalone1999, Fraschetti2011} from open field lines. 
We have herein simulated the transport of test-particle protons in the GeV-TeV energy range within a turbulent magnetic arcade with scale height $\sim 20$ Mm. Systematically varying the turbulence strength allows us to investigate the trapping effect. Figure \ref{fig:open_field} provides a schematic representation of the trajectories of two distinct energies GCRs (highlighted in red and purple), migrating from open to closed field lines.
The prolonged confinement in the magnetic arcades enhances the column density, namely likelihood of GCR interactions, thereby contributing to gamma-ray production.
Recent remote observations from Solar Orbiter (SolO), as detailed in \cite{Antolin.etal:23}, help constrain length scale and filling factor of the loop structures.


The structure of the paper unfolds as follow.
Section \ref{sec:model} provides a comprehensive description of the theoretical model, encompassing the equations defining the magnetic arcades model and the turbulent component, as well as detailing the particle motion and the proton-proton interaction time. 
The numerical setup and the methodology employed for the analysis are expounded upon in Section \ref{sec:numerical}.
Section \ref{sec:interaction_turbulence} presents the computational results. 
In particular, Section \ref{sec:location} and \ref{sec:num_int} delves into the dependence of the number and position of interacting protons on the turbulence strength and on geometrical parameters.
The direction of outgoing photons is shown in Section \ref{sec:angular}, while the comparison with the observed gamma-ray flux is presented in Section \ref{sec:flux}.
Lastly, Section \ref{sec:summary} offers a concise summary of the findings and insights garnered in this study.

\section{Theoretical model}
\label{sec:model}

\subsection{Magnetic arcades model}
 
The magnetic field model employed in this study is adapted from formulation presented in \cite{Rial2013} \citep[and originally proposed by][]{Oliver1993}.
This model comprises a potential-field arcade situated in the $xy-$plane, defined by the vector potential
\begin{equation}
    A_z (x,y) = B_0 \Lambda_B \cos{\left(\frac{x}{\Lambda_B}\right)}e^{-\frac{y}{\Lambda_B}}, 
\label{eq:Az}
\end{equation}
while $A_x (x,y) = A_y (x,y) = 0$. Here, $B_0 = 100$ Gauss denotes the magnetic field strength at the bottom of the photosphere, where the arcade is anchored,
while $\Lambda_B = L/\pi$ represents the magnetic scale height, related to the linear size of the domain $L$. 
For bottom of the photosphere ($y=0$), the surface of vertical optical depth for the $500$ nm radiation equal to unity is adopted \citep{Athay:76}.
The corresponding non-zero magnetic field components ($\bf{B} = \nabla \times A$) are given by
\begin{equation}
\begin{aligned}
\label{eq:magfield}
    B_x (x,y) &= - B_0  \cos{\left(\frac{x}{\Lambda_B}\right)}e^{-\frac{y}{\Lambda_B}} \\
    B_y (x,y) &=  B_0  \sin{\left(\frac{x}{\Lambda_B}\right)}e^{-\frac{y}{\Lambda_B}}. 
\end{aligned}
\end{equation}
Thus, the spatial variation of the magnetic field magnitude along the $z-$direction is neglected.
Figure \ref{fig:open_field} illustrates the 2D shape of the magnetic arcades (in black) and the blue color-scaled magnetic field magnitude.
Other more complex geometries (also in 3D) have been used for arcade modeling in the literature \citep[see, e.g.,][]{Zhao2017}.
However, the geometry experienced by GeV-TeV protons in the location relevant to gamma-ray production, as elaborated in Section \ref{sec:interaction_turbulence}, is 2D with a good approximation.

MHD resistive radiative simulations \citep[e.g.,][]{Gonzales2021} show that these closed structures are adjacent to macrospiculae, i.e., large spike-like jets in the solar limb \citep[][]{Gonzales2021} first identified by \cite{Bohlin1975}. Macrospiculae rise up to $70$ Mm above the solar limb in polar coronal holes, with a lifetime ranging from about $3$ up to $45$ minutes \citep{Loboda2019}. 
Adjacent closed structures have comparable size and lifetime \citep{Gonzales2021} and are assumed to be static on the time scale of GCRs transport. 
The value of $B_0$ on the photospheric surface is obtained by averaging by eye the field reconstructed from 3D simulations by \cite{Gonzales2018} over a $\sim 6^2$ Mm$^2$ region at the bottom of a macrospiculae.
It is important to note that we neglect the variation of the magnetic field strength on the surface $y=0$, although MHD simulations suggest a gradient towards the footpoints \citep{Gonzales2018}, because this gradient only affect the gyrophase of the GCRs lost into the Sun surface, not relevant to gamma-ray emission.

This investigation focuses on assessing the effect of closed magnetic field geometry on GCR proton-atmosphere proton collisions and their impact on the observed gamma-ray flux. 
The chosen geometry is supported by SolO observations of closed magnetic structures in the chromosphere and low corona \citep[see][]{Antolin.etal:23}: while some of these loops extend up to $40$ Mm, a significant fraction is limited to $10-20$ Mm. 
We note that while we use a magnetic-arcade model for this study, our results are generally also qualitatively applicable to other closed-loop configurations \cite[as the ones reported in][]{Antolin.etal:23}. Such configurations might be ubiquitous on the solar disk for sufficiently long exposure.
In support of this assumption, \cite{Linden2022} (Fig. 4) show a three years time integrated gamma-ray energy spectrum in the range $0.1-10$ GeV unchanged over the solar cycle.

\subsection{Turbulent magnetic field}
We introduce a turbulent component to the static magnetic field, following \cite{Giacalone1999}, thus defining the total magnetic field as $\mathbf{B}_\mathrm{tot}(x,y,z) = \mathbf{B}(x,y) + \delta \mathbf{B}(x,y,z)$, where $\mathbf{B}(x,y)$ represents the static magnetic field defined in Equation (\ref{eq:magfield}), while $\delta \mathbf{B}(x,y,z)$ is the turbulent component, which is independent on time and varies in space.
At every point in physical space where the particle is located, the procedure for constructing turbulence involves generating a given number $N_m$ of transverse waves, each with random values assigned for its amplitude, phase, and orientation \citep{Fraschetti2012}.
The ratio of the wave variance  to the mean square field strength of the arcade is defined as $\sigma^2= \langle \delta\mathbf{B}^2(x,y,z)\rangle/|\mathbf{B}(x,y)|^2$, where $\langle \cdot \rangle$ indicates the average over an ensemble of turbulence realizations. Note that ${\bf B}(x,y)$ is the static arcade model part, and is the same for every realization. Thus, $\sigma^2$ is taken to be a constant.
Presently, remote observations lack the capability to constrain the geometry of magnetic fluctuations in the solar chromosphere \citep[][]{Criscuoli2021}, while the amplitude could be constrained down to the photosphere using the Zeeman-Doppler measurements. Accessing the magnetic turbulence strength and inertial range in the chromosphere and high-photosphere, where gamma-ray production is likely most prominent, currently relies on extrapolation between Zeeman Doppler measurements of the radial magnetic field at the photosphere, such as those from DKIST \citep[see, e.g.,][]{Rimmele2020, Rast2021} and observations from EIS/HINODE in the low corona \citep{Culhane2007}.

\begin{figure}
    \centering
    \includegraphics[width=0.5\textwidth]{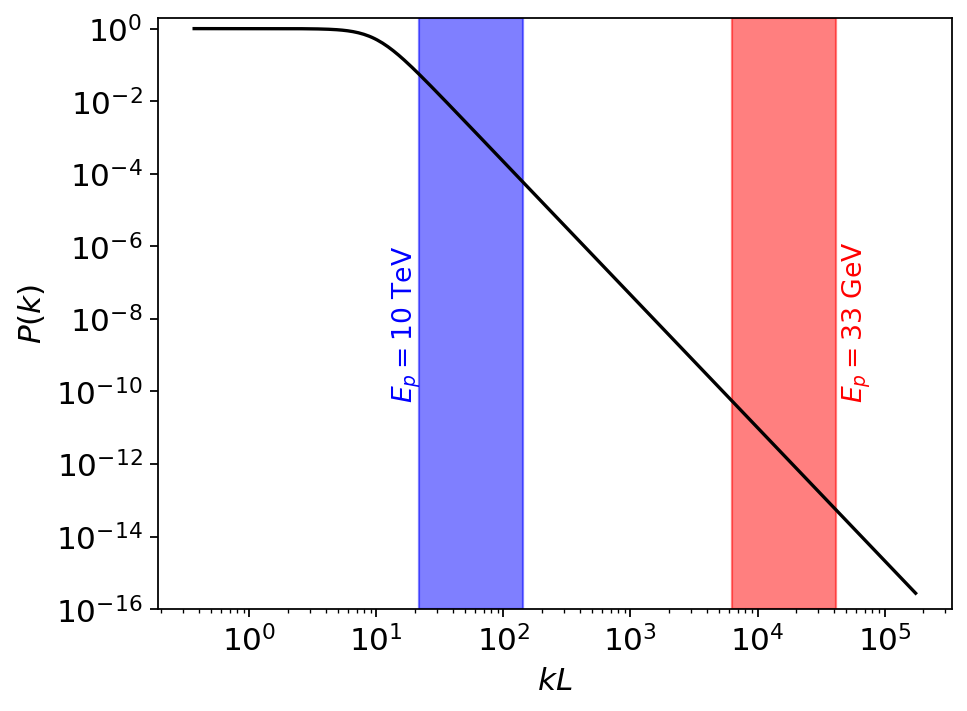}
    \caption{Power spectrum $P(k) = [1.0 + (kL_c)^q]^{-1}$ of random magnetic fluctuations, with $q=11/3$ for a 3D isotropic turbulence in all of our simulations. The blue and red ribbons represent the gyroradius calculated at the maximum ($E_p^{max} = 10$ TeV) and minimum ($E_p^{min} = 33$ GeV) energy throughout the computational box, respectively.}
    \label{fig:power_spectra}
\end{figure}

The power spectrum used to generate the magnetic fluctuations is depicted in Figure \ref{fig:power_spectra}.
Here, $L = 0.03  R_\odot$ (with solar radius $R_\odot = R_\mathrm{sun} = 6.957 \times 10^{10}$ cm), and consequently, $L_c = 0.003 \, R_\odot$, corresponding to the gyroradius of a proton with an energy of approximately $20$ TeV when injected at minimum injection altitude.
Additionally, the gyroradius for the maximum ($E_p = 10$ TeV) and minimum ($E_p = 33$ GeV) proton energy throughout the computational box is illustrated. 
In this study, we select $L_\mathrm{min} = 10^{-6} \ R_\odot$ and $L_\mathrm{max} = 5 \times 10^{5}L_\mathrm{min}$, which are respectively linked to the turbulence wavenumbers $k_\mathrm{max} = 2 \pi/L_\mathrm{min}$ and $k_\mathrm{min} = 2 \pi/L_\mathrm{max}$. 
The correlation length is set to $L_c = 0.1L$ and is assumed to be independent on height within the photospheric/chromospheric arcade.
The turbulence coefficients are calculated at the beginning of the simulation and reshuffled every 100 particles. Different values for $\sigma^2$ are chosen, namely $10^{-3}$, $10^{-1}$, and $1$.

The power spectrum of velocity fluctuation measured by CoMP, discussed in \cite{Morton2015}, covers a narrower range compared to the power spectrum considered here, which span four decades of magnetic fluctuations. This range was chosen to ensure resonant interaction across all GCR energies within an exponentially decreasing magnetic field.
By adjusting the values of $L_\mathrm{min, max}$ and $L_c$, the inertial range can be constrained to fewer than four decades leading to a suppressed diffusion, especially for protons at relatively very low ($\sim $ GeV) or very high ($\sim 10$ TeV) energy.
This adjustment alters the resonant condition, causing turbulence to affect particles of varying energies differently, thereby making direct comparison of high and low energy protons impossible due to the dissimilarities induced by the altered turbulence dynamics.


\subsection{Particle motion}
The particle speed is constant in the assumed magnetostatic field: particles travel much faster than the flow speed $\mathbf{U}$ so that the motional electric field $\mathbf{E} = - \mathbf{U} \times \mathbf{B}$ is negligible. Protons move on the total background magnetic field according to the Lorentz equation
\begin{equation}
\label{eq:particles}
 \begin{dcases} 
  \frac{d \textbf{x}}{d t} = \textbf{v} \\ 
  \frac{d (\gamma \textbf{v})}{d t} = \left(\frac{e}{mc}\right) \textbf{v} \times \textbf{B} \,.
  \end{dcases}
\end{equation}
Here, $\textbf{x}$ and $\textbf{v}$ denote the spatial and velocity coordinates, respectively,  $\gamma= (1 - \textbf{v}^2/c^2)^{-1/2}$, where $c$ is the speed of light, and $(e/m)$ is the proton charge to mass ratio.

In the arcade magnetic field described by Equation (\ref{eq:magfield}), particles experience gradient and curvature drifts induced by the inhomogeneity of the unperturbed B-field, and the corresponding guiding center velocity is given by
\begin{equation}
\label{eq:drift}
    \mathbf{v}_{GC} = \mathbf{v}_G + \mathbf{v}_C = \frac{\gamma m_pc B_0^2}{2eB^3\Lambda_B}(v_\perp^2 + 2v_\parallel^2) \cos{\left(\frac{x}{\Lambda_B}\right)}e^{-\frac{2y}{\Lambda_B}} \hat{\mathbf{k}},
\end{equation}
where $v_\perp$ and $v_\parallel$ represent the particle velocities perpendicular and parallel to the magnetic field, respectively. 
Although $\mathbf{v}_{GC}$ aligns parallel to the z-axis in the unperturbed magnetic field (see Equation \ref{eq:magfield}), particles predominantly undergo in-plane drift since $\mathbf{v}_{GC}/v \simeq r_g/\Lambda_B$, where $r_g$ represents the particle gyroradius. 

Since particles can move along the $z$-direction as well, it is essential to impose a limit on the particles displacement along the $z$-axis: particles are considered escaped (hence, no gamma-ray emission) when they exit the computational domain in the $x-y$ plane and/or $z_p <  -0.01 \ R_\odot$ or $z_p > 0.01 \ R_\odot$.
Escape can be caused by drifts or cross-field diffusion. Escape can occur inside the solar photosphere ($y<0$), or outside the box, with plasma too rarefied to induce gamma-ray production.

\subsection{GCR-proton collision}
\label{sec:cross_section}
In the context of gamma-rays produced by proton-proton collisions, it is essential to consider the interaction time $t_\mathrm{int}$ of these collisions. 
The gamma-rays are produced in the highly dense low chromosphere and photosphere plasmas.
The ratio between the total time elapsed from the initial time $t=0$ to the final time $t_f$ (i.e., $\Delta t$) and the interaction time is expressed as
\begin{equation}
\label{eq:t_int}
    \Delta t / t_\mathrm{int} = \int_{t_i}^{t_f} dt/t_{int}(y) =   \sigma_\mathrm{pp} v \int_{t_i}^{t_f} n[y(t)] dt.
\end{equation}
Here, $dt$ is the code time-step, and $n[y(t)]$ represents the solar atmosphere number density, assumed uniform during each time step, that is a fixed value for all particle energies.
The average particle speed is denoted by $v \simeq c$, and $\sigma_\mathrm{pp}$ represents the total proton-proton inelastic collision cross section \citep[see, e.g.,][]{Tanabashi2018}.
\cite{Kafexhiu2014} computed $\sigma_\mathrm{pp}$ and parameterized it as follows
\begin{equation}
\begin{aligned}
\label{eq:cross_section}
    \sigma_\mathrm{pp} (\gamma) =& \left[30.7 - 0.96\log\left(\frac{T}{T_{th}}\right) +0.18\log^2\left(\frac{T}{T_{th}}\right)\right] \\  & \times \left[1 - \left(\frac{T_{th}}{T}\right)^{1.9}\right]^3 \ \mathrm{mb},
\end{aligned}
\end{equation}
where $T = m c^2(\gamma - 1)$ is the proton kinetic energy in the laboratory frame, while $T_{th} = m c^2(\gamma^\mathrm{th} - 1) \approx 0.2797$ GeV is the threshold kinetic energy for the production of neutral and charged pions.
This parameterization is based on the best fit to measurements published by the TOTEM Collaboration at the LHC \citep[see, e.g.,][]{Antchev2013A, TOTEM2013}, as reported in \cite{Kafexhiu2014}.
Due to momentum conservation, photons preserve the direction of the outgoing GCRs at the instant of interaction.
The energy of the outgoing gamma-rays is approximately ten times lower than the initial GCR proton energy \citep[$E_\gamma \sim 0.1E_p$,][]{Kelner2006}.

\section{Numerical setup}
\label{sec:numerical}
The numerical simulations are conducted using the PLUTO code originally designed for astrophysical gas dynamics \citep[see][]{Mignone2007, Mignone2012}.
The grid resolution is configured as $1000 \times 1000$, with the $j$ index representing the number of cell along the $y$-direction.
We opted to model a simplified 2D magnetic arcade rather than employing pre-existing MHD grids  \citep[such as MHD simulations of solar magnetic atmosphere obtained with the MURaM code, see][]{Vogler2005} as it is significantly less computationally demanding.


\subsection{Computational box size}
\label{sec:box_size}
The initial configuration involves a 2D computational domain with dimensions $L \times L$, where $L = 0.03 R_\odot$.  
The computational domain is specifically defined within $-0.015 < x/R_\odot < 0.015$ and $0 < y/R_\odot < 0.03$.
This chosen length scale for the loop arcades aligns with recent observations of Solar Orbiter \citep[][]{Antolin.etal:23} and numerical simulations by \cite{Bale2021} and \cite{Wyper2017}.
A lower limit on the magnetic arcades lengthscale is provided by the high-resolution observations of the New Vacuum Solar Telescope and Solar Dynamics Observatory \citep{Duan.etal:23}.
These observations have also reported evidence of magnetic reconnection in bright patches with a transverse size $5^{''} \times 5^{''}$ ($\sim 3,000$ km $\times$ $3,000$ km) between macrospiculae jets and adjacent closed loops, that the length scale of our simulations are consistent with.

The vertical profile of the plasma density in the low photosphere is still debated \citep{MacBride2022,Morton2023,Yalim.etal:23}, and hardly accessible to observations. We have tested a variety of numerical configurations and profiles: placing the lower boundary of the simulation box ($y=0$ plane, see Fig.\ref{fig:open_field}) both at the bottom (a) and top (b) of the photosphere.
(a) 
At the bottom of the photosphere, the mass density is estimated as $\rho_0 \sim 10^{-6} \ g/cm^3$ \citep[][]{Morton2023}, resulting in a proton (or equivalently neutral hydrogen atoms) number density $n_0 = \rho_0/m = 10^{18}$ $cm^{-3}$.
(b) The top of the photosphere (or low chromosphere), can be located at a height of $\sim 500$ km \citep[see, e.g.,][]{MacBride2022}  with a mass density of $\rho_0 \sim 10^{-8} \ g/cm^3$ \citep{Morton2023}. 

\cite{Gonzales2018} also proposed a solar atmospheric model \cite[extended up to 30 Mm in][]{Gonzales2021} characterized by a lower density ($\rho_0 \sim 10^{-10} \ g/cm^3$) at the bottom of the photosphere ($y = 0$ in Fig. \ref{fig:open_field}) compared to \cite{Morton2023}. This discrepancy arises from \cite{Gonzales2018} utilization of a solar atmospheric model derived from the C7 model outlined in \cite{Avrett2008}, coupled with a 3D potential magnetic field configuration extrapolated from a realistic photospheric quiet-Sun model. The C7 model, rooted in the SUMER atlas of the extreme ultraviolet spectrum \citep{Curdt1999}, exhibits notable consistency with both line intensities and profiles. Conversely, the solar atmospheric model detailed in \cite{Morton2023} is derived from a sequence of models by \cite{Fontenla1993}, wherein only the quiet Sun profile is considered.
In the remainder of the paper we present results for case (a) only, which incorporates the density profile from \cite{Morton2023}, since the low chromospheric density is not sufficient to explain the gamma-ray spectrum.


\subsection{Particles initialization and orbit integration}
Test particle protons are initialized with one particle per cell at discrete heights from the surface $y=0$ labelled by the $j$ index and representing the height of migration of GCRs from open flux tubes to the arcades (see Fig. \ref{fig:open_field}). 

\begin{figure}
    \centering
    \includegraphics[width=0.48\textwidth]{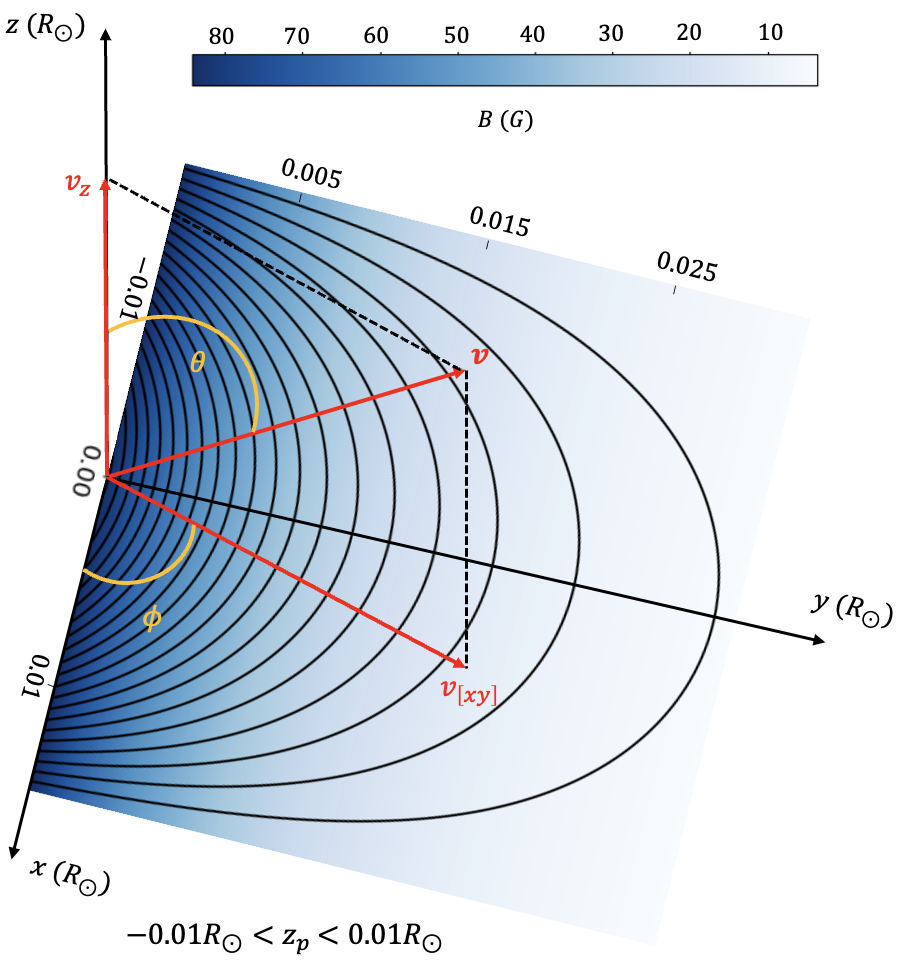}
    \caption{Schematic decomposition of the particle velocity vector in its components and angular coordinates $\theta$ and $\phi$ over the magnetic arcade domain already described in Figure \ref{fig:open_field}.}
    \label{fig:arcade_3D}
\end{figure}

Particle injection is performed within the specific ranges $200 < j < 210$,  $400 < j < 410$,  $600 < j < 610$, and $800 < j < 810$.
In each injection case, there are a total of $9,000$ test particle protons at the initial time.
The initial velocity distribution is assumed to be isotropic: 
\begin{equation}
\label{eq:velocities}
 \begin{dcases} 
  v_x = v_0 \sin \theta \cos \phi \\ 
  v_y = v_0 \sin \theta \sin \phi \\ 
  v_z = v_0 \cos \theta, 
  \end{dcases}
\end{equation}
where $\theta = \arccos(1 - 2 \mathcal{R}_{[0,1]})$ and $\phi = 2 \pi \mathcal{R}_{[0,1]}$ (see Figure \ref{fig:arcade_3D}).
In these expressions $\mathcal{R}_{[0,1]}$ represents a random number between 0 and 1. 
An isotropic velocity accounts for the protons entering the arcade from the top, those grazing the Sun surface and those back scattering inside the flux tube before entering the arcade in their motion upward. 


We employ the Boris algorithm, already implemented in the PLUTO code, to solve 
Equation \ref{eq:particles}. 
The integration is performed until the end of the computational time defined as  the time of escape or interaction of the last GCR ($t_\mathrm{stop} = 20 \ R_\odot/c \simeq 46.4 \ s$).
The Boris integrator is chosen as the fiducial scheme due to its time reversibility and good conservation properties for long-time simulations.

The magnetic field $\mathbf{B}$ of the arcade plus the turbulent component is properly interpolated at the particle position, following the approach of \cite{Mignone2018}.
The interpolated value at the particle position of any grid quantity (such as the magnetic field) is determined through a weighted sum achieved with a conventional field weighting approach \citep[see][]{Birdsall1991}. In practical terms, each particle's contribution is significant only within its corresponding computational zone, along with its adjacent left and right neighbors \citep[see][for further details]{Mignone2018}.

\subsection{Energy dependence of cross-field diffusion coefficient}
\label{sec:diff_coeff}
A closed arcade can contribute to the gamma-ray flux only if cross-field diffusion occurs.
In the absence of turbulence ($\sigma^2 = 0$), particles follow individual field lines, as documented in previous studies \citep[see, e.g.,][]{Jokipii1993, Giacalone1999}, and may undergo mirroring, as postulated by SSG91.
Conversely, when turbulence is present ($\sigma^2 \ne 0$), particles also experience both pitch-angle scattering and cross field diffusion, which enable them to cross magnetic field lines. These mechanisms are expected to enhance the residence time of particles within the dense atmosphere.
The GCRs transport from open flux tubes to adjacent closed magnetic field arcades shown in Figure \ref{fig:open_field} occurs via cross-field diffusion perpendicular to the open flux tube magnetic field within the arcade plane. Out-of-plane cross-field diffusion is not relevant here, as it directs particles towards distinct magnetic arcades or open flux tubes.

The diffusion is described by coefficients, namely $\kappa_\parallel$ and $\kappa_\perp$, aligned with and perpendicular to the average magnetic field, respectively.
Vast is the literature on the calculation of the diffusion coefficients \citep[see, e.g.,][]{Parker1965, Chapman1970, Forman1977, Bieber1997, Giacalone1999}.
It is essential to note that the perpendicular diffusion addressed in $\kappa_\perp$ in this work pertains to cross-field diffusion, distinct from the diffusion perpendicular to the average magnetic field calculated in, e.g., \cite{Giacalone1999}.

The diffusion in the SSG91 model was affected by radial distance and energy with an empirical power-law index. 
Here, we make use of the dependence of the cross-field diffusion for 3D isotropic turbulence both on the turbulence strength $\sigma^2$ and the particle energy \citep[see, e.g.,][]{Giacalone1999, Fraschetti2011}. 


The main ansatz of this work is that the observed flatness of the Fermi-LAT gamma-ray flux for $E_\gamma < 30 - 100$ GeV stems from the increase with energy of the cross-field diffusion coefficient.
Specifically, $\kappa_\perp$ is smaller for low energy particles, with a resulting suppressed migration from open to closed field lines compared to high-energy particles.
Thus, if $\kappa_\perp = 0$, low-energy particles precipitate into the photosphere, mirror within the flux tube and emit a gamma-ray before reaching the chromosphere during their outward motion \citep{Li2024}. 
Such a flux is not included in the present calculation.  
Therefore, the inclusion of the contribution from flux tubes may result in a higher intensity of the calculated gamma-ray flux. Notably, the gamma-ray flux obtained by \cite{Li2024} appears slightly lower than the observed flux, indicating the necessity to account for both open and closed magnetic field lines.

The expression of $\kappa_\perp$ due to the cross-field diffusion in the 3D isotropic turbulence in terms of $\sigma^2$ and proton energy is provided by \cite{Fraschetti2011}:
\begin{equation}
\label{eq:perp}
    \frac{\kappa_\perp}{\kappa_B} (E_p) = \frac{\pi}{40} \sigma^2 \frac{q-1}{q-2}r_g(E_p) k_\mathrm{min},
\end{equation}
where $\kappa_B = (1/3)r_g(E_p)c$ denotes the Bohm diffusion coefficient, and $k_\mathrm{min} = 2\pi/L_c$.
It should be noted that the condition $k^\mathrm{min}_\parallel v_\parallel t \ll 1$ implying $k^\mathrm{min}_\parallel \approx k_\mathrm{min}$  (where $k_\mathrm{min} \sim 2094/R_{\odot}$) for the validity of Equation (\ref{eq:perp}), is not strictly satisfied if the particles free-stream.

By using Equation (\ref{eq:perp}), the average square displacement for GCRs migrating into the magnetic arcade can be estimated as $\sqrt{\langle \Delta x^2 \rangle}\lesssim \sqrt{2 \, \Delta t \, \kappa_\perp (E_p)}$. 
Here, $\Delta t$ is the transit time for a scatter-free GCR propagating to $y = 0$ in the flux tube adjacent to the magnetic arcade. For a generic height $y$ within the arcade, and with $y^* \simeq L =0.03R_\odot$ being the height of the magnetic arcade, and using $E_p = 100$ GeV and a magnetic field magnitude $B = B_0 e^{- y/\Lambda_B} = 100$ G (for $y= 0$), the resulting expression is
\begin{equation}
   \resizebox{0.47\textwidth}{!}{$ \frac{\sqrt{<\Delta x^2>}}{R_\odot}(y) \approx 4 \times 10^{-4}  \sqrt{\sigma^2 \left(\frac{y^* - y}{c}\right)} \left(\frac{E_p}{100  \,\mathrm{GeV}}\right) \left(\frac{e^{y/\Lambda_B}}{B_0/100  \,\mathrm{G}}\right)$}
    \label{eq:deltax}
\end{equation}
The $\sigma^2$-dependence in Equation (\ref{eq:deltax}) suggests that GCR with energies below a certain threshold ($E_p = E_\mathrm{thr} \sim 10^2-10^3$ GeV for $\sigma^2 = 1$ and $E_\mathrm{thr} \sim 10^3-10^4$ GeV for $\sigma^2 = 0.001$) cannot efficiently diffuse across field lines and migrate from open flux tube into the arcades. 

The scaling of $\kappa_\parallel \sim E_p^{1/3}$ for relativistic particles leads to
\begin{equation}
    \frac{\kappa_\parallel}{\kappa_B}(E_p) \sim \frac{1}{\sigma^2} E_p^{-2/3}.
\end{equation}
We note that above the energy scaling of $\kappa_\parallel$  for 3D isotropic turbulence has been used even for $\sigma^2 = 1$, which is very close to the Bohm regime. 
\cite{Fraschetti2012} show that this assumption is reasonable for $\sigma^2 \simeq 1$.

Since GCRs migrate into the arcade only during the time that the parallel diffusion can confine them within the adjacent open flux tube, the gamma-ray flux needs a correction proportional to the relative displacements as given by
\begin{equation}
    \resizebox{0.48\textwidth}{!}{$\xi = \left(\frac{<\Delta x^2>}{<\Delta y^2>} (E_p) \right)^{1/2} = \begin{dcases} 
  \left(\frac{\kappa_\perp/\kappa_B}{\kappa_\parallel/\kappa_B} (E_p)\right)^{1/2} &\text{for $E_p < E_\mathrm{thr}$} \\ 
 1 &\text{for $E_p > E_\mathrm{thr}$}\,
  \end{dcases}$}
    \label{eq:coeff_ratio}
\end{equation}
If $\kappa_\perp/\kappa_\parallel \sim 1$, the migration into the arcade is very efficient and the GCR flux into the arcade is comparable with the GCR flux impinging on the Sun. If $\kappa_\perp/\kappa_\parallel \ll 1$ most GCR travel close to free-streaming regime into the open flux tube, with a resulting suppression of the GCRs flux into the arcade, hence of the gamma-ray flux. 
Therefore, the correction factor in Equation (\ref{eq:coeff_ratio}) scales (for $E_p < E_\mathrm{thr}$) with energy and turbulence strength according to 
\begin{equation}
\label{eq:propto}
    \xi \propto E_p^{5/6} \sigma^2,
\end{equation}
with a strong dependence on the turbulence relative power $\sigma^2$.

\subsection{Protons of interest and gamma-ray flux estimation}
We restrict to outgoing protons ($v_y > 0$) that produce gamma-rays observable at Earth via the collision $p+p \rightarrow \pi_0 \rightarrow \gamma + \gamma$ and to interaction occurring within the arcade ($\Delta t/t_\mathrm{int} > 1$), as interactions in the low-corona are unlikely due to the rapid decrease of the plasma density.  

The gamma-ray flux at a given photon energy $E_\gamma$ is calculated using the expression from \cite{Kelner2006}, as done also in \cite{Li2024}:
\begin{equation}
\label{eq:gamma_flux}
    \resizebox{0.47\textwidth}{!}{$\Phi^*_\gamma(E_\gamma) \equiv \frac{dN_\gamma}{dE_\gamma} = c n_p \int_{E_\gamma}^\infty \sigma_\mathrm{pp}(E_p) J_p(E_p)F_\gamma \left(\frac{E_\gamma}{E_p}, E_p\right)\frac{dE_p}{E_p}$},
\end{equation}
where $n_p$ is the density \citep[from][extended to higher altitudes by \citealp{Gonzales2021}]{Morton2023} integrated along the $y$-direction, $\sigma_\mathrm{pp}(E_p)$ is the inelastic cross section of the proton-proton collision in Equation (\ref{eq:cross_section}), $F_\gamma(E_\gamma/E_p, E_p)$ is calculated in Equation 58 by \cite{Kelner2006} and takes into account the number of photons produced per energy interval per collision. The proton auxiliary function $J_p(E_p)$ is defined as 
\begin{equation}
\label{eq:Jp}
    Jp(E_p) = \xi \frac{N_\mathrm{int}}{N_\mathrm{inj}}(E_p) \left(E_p^2 \frac{dN}{dE_p}\right)_\mathrm{obs}\times \frac{1}{E_p^2 c},
\end{equation}
in units of $N_\mathrm{protons} \cdot$ GeV$^{-1} \cdot cm^{-3}$.
In this expression, $N_\mathrm{int}/N_\mathrm{inj}$ is the ratio of the number of protons interacting within the arcade to the total number of injected particles ($N_\mathrm{inj} = 9,000$). 
Although $N_\mathrm{int}/N_\mathrm{inj}$ is a function of $E_p$, integrating $\Phi_\gamma(E_\gamma)$ over three decades in $E_p$ around the peak of $F_\gamma(E_\gamma/E_p, E_p)$ as a function of $E_\gamma/E_p$ produces the same results as taking its value at $E_p$, due to the rapid decrease of $F_\gamma$ around the peak.
The fraction $N_\mathrm{int}/N_\mathrm{inj}$ is also averaged over the injection altitude $j$, as well as the gyroradius for the correction $\xi$ in Equation (\ref{eq:perp}), due to the B-field dependence on the injection altitude.

For the incoming (approximately) isotropic proton GCR energy spectrum $(E_p^2 dN/dE_p)_\mathrm{obs}$ we have used the power-law fit (with power-law index $-0.7$) of the observed differential spectrum collected by the Alpha Magnetic Spectrometer (AMS) as documented in \cite{Aguilar2021} for energies $E_p \lesssim 1$ TeV. 
At higher energies, we considered the measurements from the ISS-CREAM Experiment \citep[see][]{Choi2022}.
We emphasize that the expressions in Equations (\ref{eq:gamma_flux}) and (\ref{eq:Jp}) are valid for energy range of primary protons of $0.1$ TeV $\leq E_p \leq 10^5$ TeV and for $E_\gamma/E_p \gtrsim 10^{-3}$; thus the integral in Equation (\ref{eq:gamma_flux}) is evaluated from $E_\gamma$ to $10^3 E_\gamma$.

%
%


The gamma-ray flux in Equation (\ref{eq:gamma_flux}) is normalized to 
\begin{equation}
    \Phi_\gamma(E_\gamma) = \frac{2\pi R_\odot^2}{(0.02 \times 0.03) R_\odot^2}  \left(\frac{R_\odot}{R_\mathrm{1AU}}\right)^2 \Phi^*_\gamma(E_\gamma),
    \label{eq:flux:final}
\end{equation}
where the first fraction on the right-hand side is the reciprocal of the filling factor (where $0.02 R_\odot$ and $0.03 R_\odot$ represent the size of the computational box in the $z$- and $x$-direction, respectively), assumed to be constant during the solar cycle \citep[consistently with][Fig. 4]{Linden2022}, while $R_\mathrm{1AU}$ is the distance from the Earth to the Sun.
We note that, although Equation (\ref{eq:gamma_flux}) also includes the $\eta$-meson decay into gamma-rays, \cite{Kelner2006} show that the channel of $\pi^0$-decay is dominant. 


\section{Results}
\label{sec:interaction_turbulence}

\subsection{Location of the proton-proton interaction}
\label{sec:location}

\begin{figure*}
    \centering
    \includegraphics[width=0.49\textwidth]{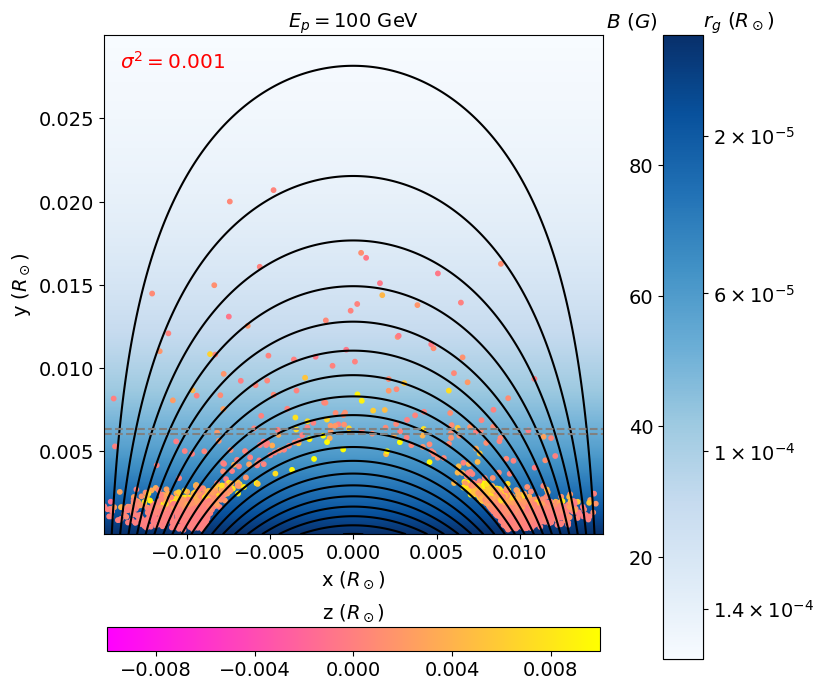}
    \includegraphics[width=0.49\textwidth]{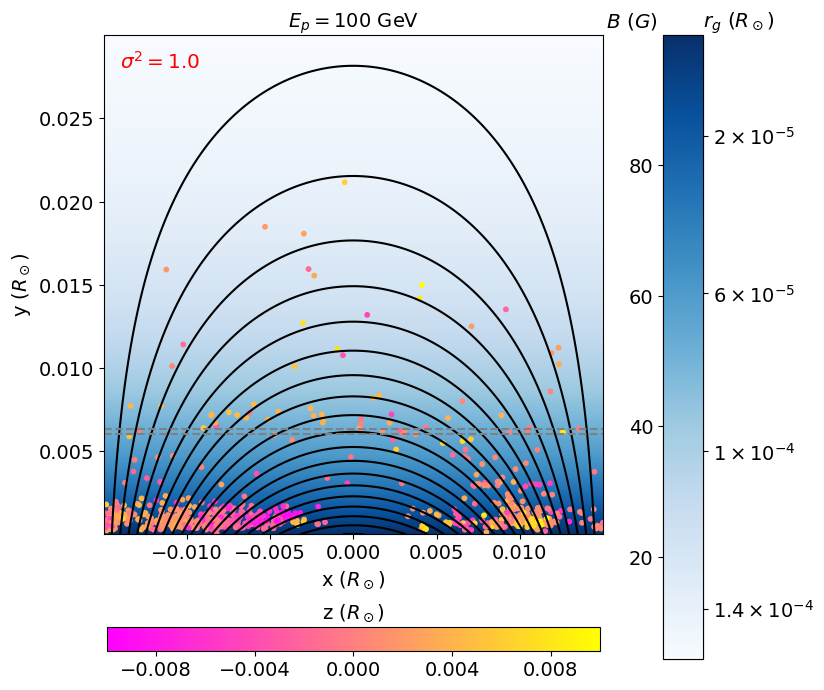}
    \includegraphics[width=0.49\textwidth]{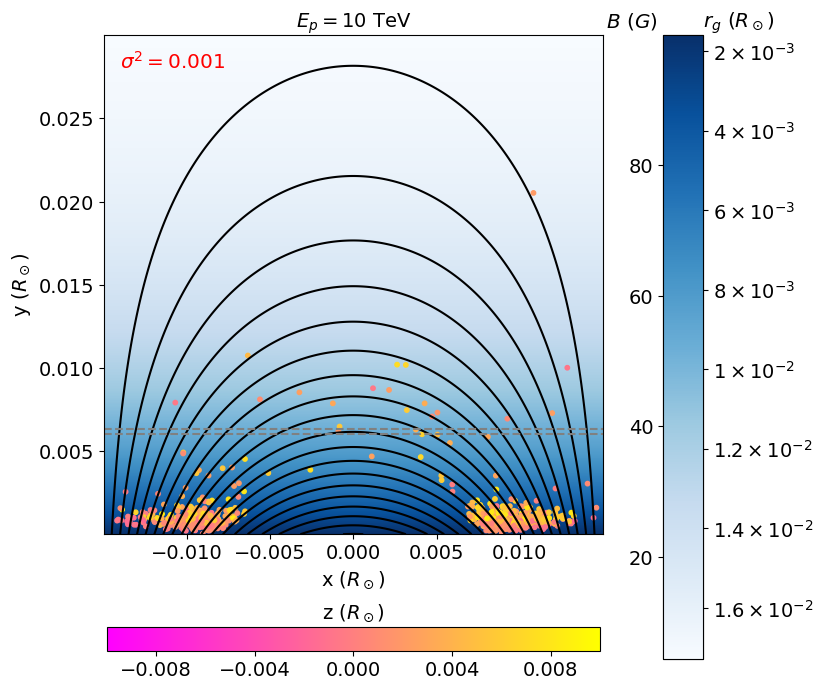}
    \includegraphics[width=0.49\textwidth]{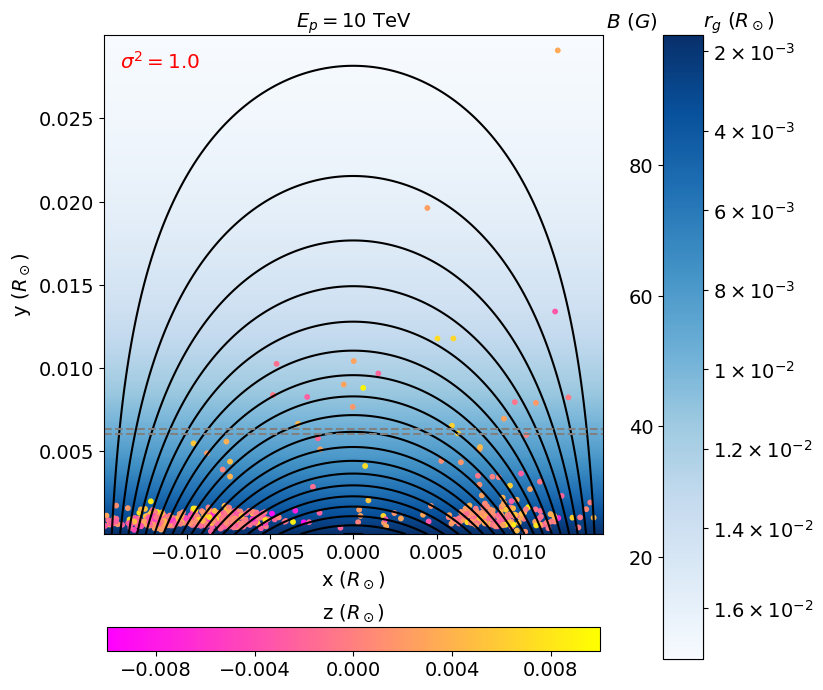}
    \caption{\textit{Top panels}: Arcade magnetic field lines (black solid) with superimposed the position at $t_\mathrm{int}$ of $100$ GeV protons over the whole computational time, color-coded in yellow-pink by their z-component obtained with $\sigma^2 = 0.001$ (left panel) and $\sigma^2 = 1$ (right panel). The blue-white color-scale represents the magnetic field magnitude and the gyroradius therein. The horizontal dashed gray lines delimit the particles injection strip.
    \textit{Bottom panels:} Same as top panels but for the $10$ TeV case.}
    \label{fig:int_y_500.png}
\end{figure*}

We examine the location of the GCR-proton interaction for distinct values of $\sigma^2$ for GCRs diffusively migrated into the arcade and undergoing collision with chromo-/photospheric protons within the arcade.
The particle injection was confined to a horizontal strip within a variety of heights; Figure \ref{fig:int_y_500.png} shows the case $200 < j < 210.$
This figure captures the positions of the interacting protons (with $v_y > 0$ and $\Delta t/t_\mathrm{int} > 1$) color-coded by their $z$-spatial coordinate at $t = t_{int}$ for two scenarios: $\sigma^2 = 0.001$ (left panel) and $\sigma^2 = 1$ (right panel).
The effect of the turbulence on the transport within the arcade is illustrated for two benchmark $E_p$ values: $100$ GeV (upper panels) and $10$ TeV (lower panels).
Horizontal dashed gray lines delineate particle position at the initial time ($t = 0$). 
The blue-white color scale represents both the magnetic field magnitude and the gyroradius.

Figure \ref{fig:int_y_500.png} confirms the expectation that most interactions occur in the low photosphere, due to very rapid plasma density vertical increase downward.
In this specific region, the gyroradius is smaller owing to the intensified magnetic field, resulting in a predominantly 2D geometry experienced by GCRs.
This finding aligns with the gamma-ray emission altitude reported by \cite{Li2024}.

In the case of weak turbulence ($\sigma^2 = 0.001$, left panels), protons just follow the magnetic field lines, undergoing mirroring.
This process is particularly efficient at $100$ GeV (upper left panel) owing to the shorter gyroperiod compared to the travel time. 
Conversely, in the scenario of strong turbulence ($\sigma^2 =1$, right panels), particle positions at the time of interaction exhibit greater dispersion along the $x$-axis at heights lower than the injection one, due to the higher cross-field diffusion in higher $\sigma^2$ environment.
This effect is especially pronounced in the $10$ TeV case (lower right panel) due to the larger gyroradius. 
Consequently, a greater number of particles interact in the lower part of the domain closer to the central axis (i.e., $x \rightarrow 0$).
This behaviour is absent in the case of weak turbulence, where particles tend to follow magnetic field lines.
Additionally, the turbulent magnetic field component along the $z$-axis contributes to spread the interaction points along the $z$-direction in the strong turbulence case. 

We note that the peak of the distribution of the interaction time occurs at $t_\mathrm{int} \simeq 0.1$ s for higher energy particles and  $t_\mathrm{int} \simeq 5$ s for lower energy particles regardless of $j$, as compared with a crossing time $t_\mathrm{cross} = L/c \sim 0.07$ s. 
The $t_\mathrm{int}$ peaks earlier for the high-energy particles, as a significant fraction of protons escape the arcade structure before interacting due to their larger gyroradius.
Nonetheless, since $t_\mathrm{int} \gg t_\mathrm{cross}$ consistently, this indicates efficient trapping of particles in the magnetic arcades.
Furthermore, we find that most GCRs interact within $5$ seconds from injection, consistently with our magnetostatic approximation.
In this context, radiative MHD simulations by \cite{Przybylski2022} demonstrate that the lifespan of closed loop structures and associated fluctuations, arising from their evolution (such as the fluctuating transverse motion of magnetic footpoints), spans on the order of a hundred seconds, surpassing the interaction time considerably. 

\subsection{GCR-proton interaction vs escape}
\label{sec:num_int}

\begin{figure*}
    \centering
    \includegraphics[width=1.0\textwidth]{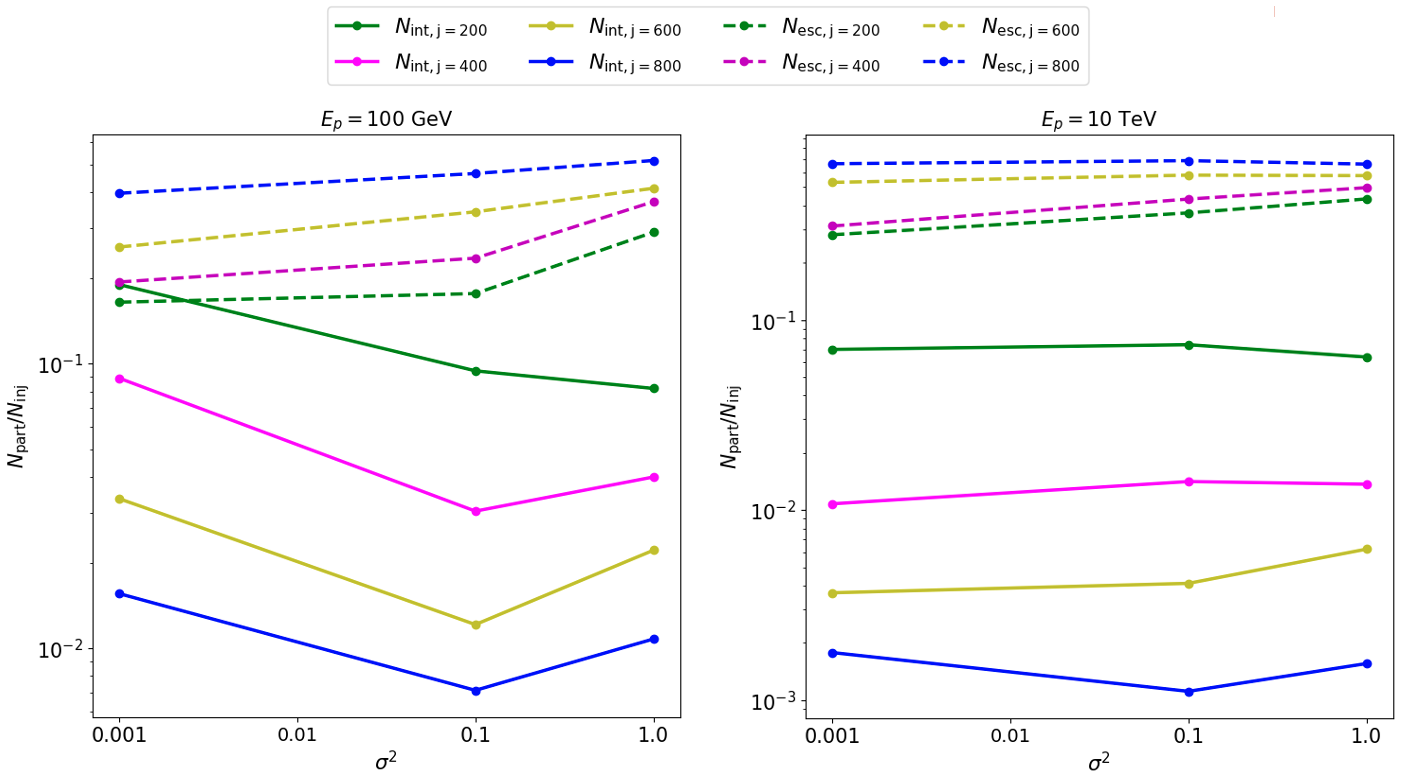}
    \caption{\textit{Left panel:} Ratios $N_{int}/N_{inj}$ (solid lines) and $N_{esc}/N_{inj}$ (dashed lines) as a function of $\sigma^2$ for different values of the injection height $j$ (from $j=200$ to $j=800$), represented by different colors, for $100$ GeV GCR protons. \textit{Right panel}: Same as left panel but for $10$ TeV GCR protons.}
    \label{fig:trend_int}
\end{figure*}

Figure \ref{fig:trend_int} illustrates the dependence on $\sigma^2$ of the number of interacting particles ($N_\mathrm{int}$, solid lines) and of GCRs escaping from the simulation box before an interaction within the arcade ($N_\mathrm{esc}$, dashed lines), both in units of $N_\mathrm{inj}$. 
The discrete injection heights $j$ are denoted by distinct colors for the two energies 100 GeV (left panel) and 10 TeV (right panel). 

Focusing on the $10$ TeV case (right panel), $N_\mathrm{int}$ (solid lines) remains relatively constant with varying $\sigma^2$ across all injection altitudes.
Since the gyroradius is large, turbulence does not have an important effect on $N_\mathrm{int}$. 
This is primarily attributed to the rapid escape of high-energy particles from dense regions, driven by their larger gyroradius.
Notably, $N_\mathrm{int}$ decreases with increasing injection altitudes $j$, as the protons counted in the green solid line ($j=200$, shorter version for $200 < j < 210$) are injected into a high-density region compared to those in the blue solid line ($j=800$, shorter version for $800 < j < 810$).
As a result, $N_\mathrm{esc}$ (dashed lines) also exhibits minimal variation, remaining approximately constant across different values of $\sigma^2$.
Notably, $N_\mathrm{esc}$ is smaller at lower altitude due to the higher likelihood of interaction in high-density plasma than escape. Conversely, at higher $j$ escape is more likely.
In general, $N_\mathrm{esc}$ at higher energy is larger since the larger gyroradius enables them to escape the magnetic structure more rapidly.

On the contrary, in the $100$ GeV case (left panel), turbulence exerts a stronger influence, shown by the decrease (increase) of $N_\mathrm{int}$ ($N_\mathrm{esc}$) with increasing $\sigma^2$. 
This counterintuitive behaviour is primarily driven by the smaller gyroradius. 
Under conditions of strong turbulence, fewer protons interact due to the increase in the cross-field diffusion $\kappa_\perp$ with $\sigma^2$, causing protons to exit the higher density region (i.e., shorter interaction time) more rapidly. This escape is due to the larger scattering mean path, $\lambda_\parallel \propto B^{1/3}$, in regions at high altitude (large $y$) with lower B-field \citep{Fraschetti2022}.
Indeed, as shown in Figure \ref{fig:int_y_500.png}, the mirroring is not efficient at keeping particles trapped within the arcade as in the low-turbulence case.
Additionally, we observe an increased tendency of escaping along the $z$-axis (as outlined in Equation \ref{eq:drift}) in the presence of strong turbulence, as protons follow the z-component of $\delta \mathbf{B}$ (the unperturbed field $B$ has no z-component, see Eq. \ref{eq:magfield}). Such an escape along the z-axis occurs in the real arcade field provided that the arcade ribbon is thin along the z-axis relative to $r_g$; if the ribbon is very thick, the arcade magnetic field can capture those GCRs in their escape along z-axis, possibly contributing to the outgoing gamma-ray flux. 
In summary, we find that the trapping effect induced by turbulence is strongly contingent on particle energy: at high energies, $N_\mathrm{int}$ remains relatively unaffected by turbulence, whereas at low energies, $N_\mathrm{int}$ decreases with increasing $\sigma^2$.

The escaping particles with both $v_y > 0$ and $v_y < 0$ are conglomerated in one single group (dashed lines) in Figure \ref{fig:trend_int}. However, upon separate analysis, we found that the number of particles escaping with $v_y > 0$ exceeds those escaping with $v_y < 0$ (about $60 \%$ vs $40 \%$). Nevertheless, it is improbable that the protons escaping from the top of the computational domain ($v_y > 0$) will interact, given the sharp decrease in the density profile with altitude \citep{Morton2023}.
The protons escaping from the sides of the computational domain (at low altitudes) with $v_y <0$ are expected to produce downward gamma-rays, which are not significant for our gamma-ray flux calculation.
In addition, we have separately counted the GCRs precipitating into the Sun ($y<0$) and found a decrease with $\sigma^2$.
Therefore, turbulence effectively prevents the downward movement of particles.
A discernible reduction is noted with the elevation of injection altitude $j$, implying that particles introduced at lower altitudes tend to precipitate downward, particularly in scenarios characterized by low $\sigma^2$, where particles predominantly follow magnetic field lines that direct them downwards.



\begin{figure*}
\centering
\includegraphics[width=0.48\textwidth]{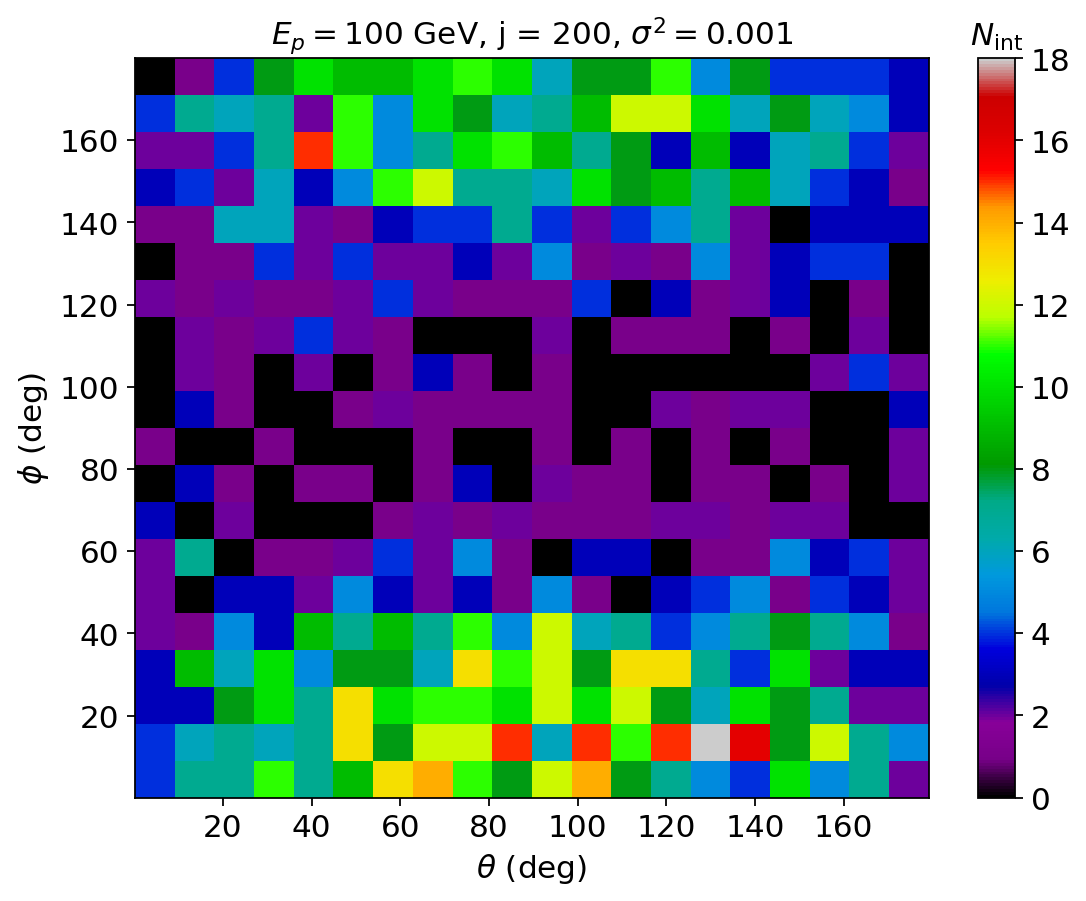}
    \includegraphics[width=0.48\textwidth]{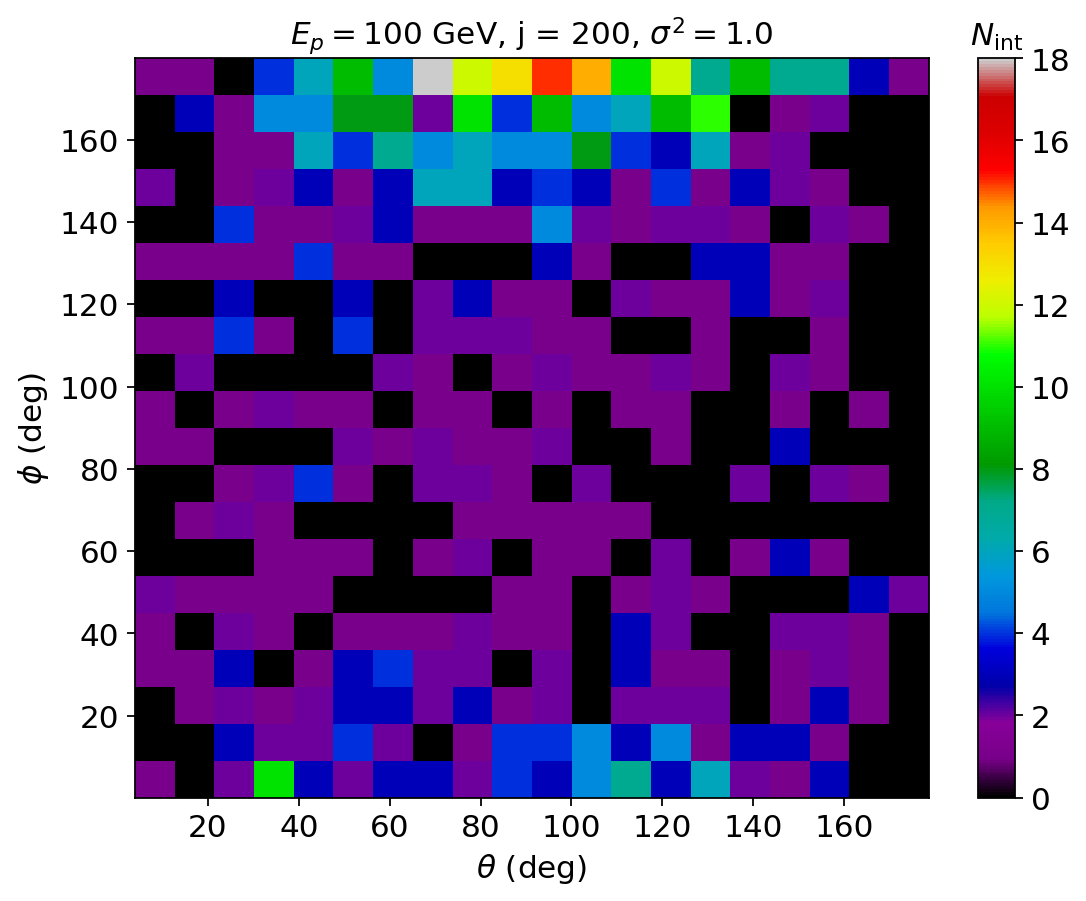} \\
\includegraphics[width=0.48\textwidth]{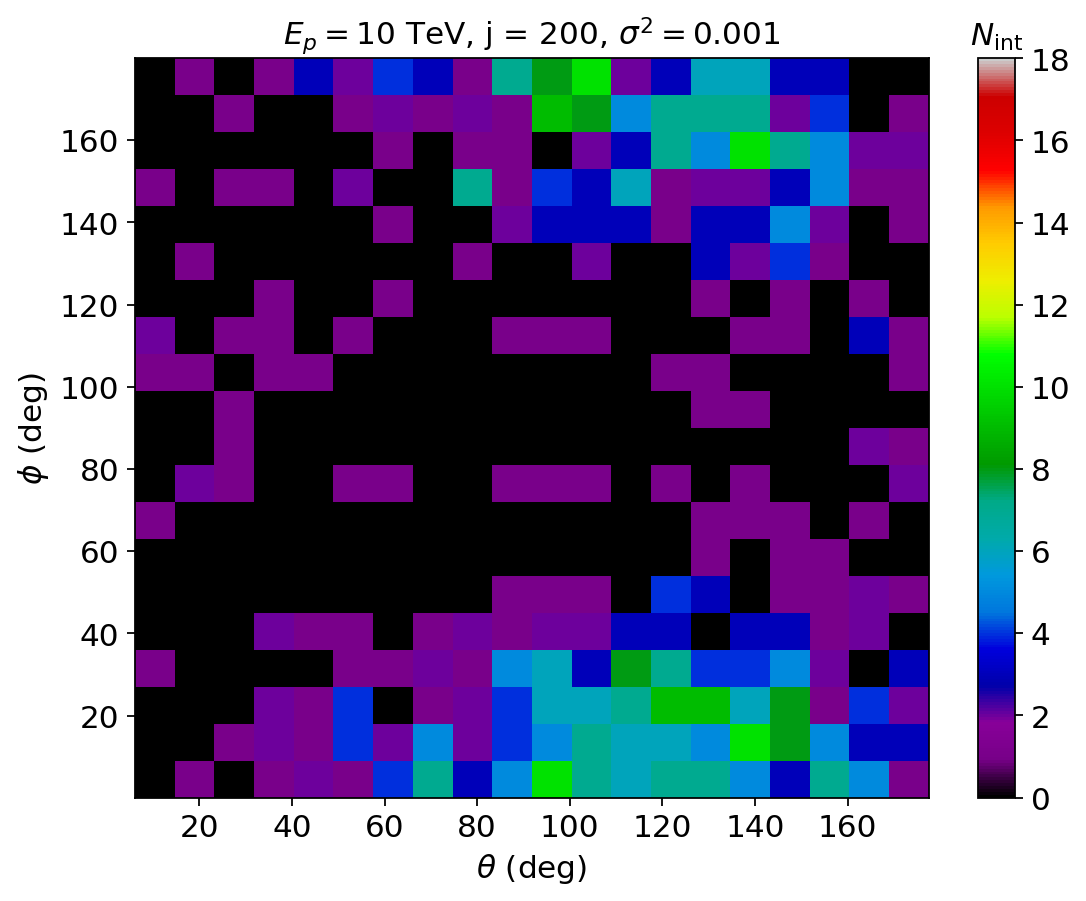}
    \includegraphics[width=0.48\textwidth]{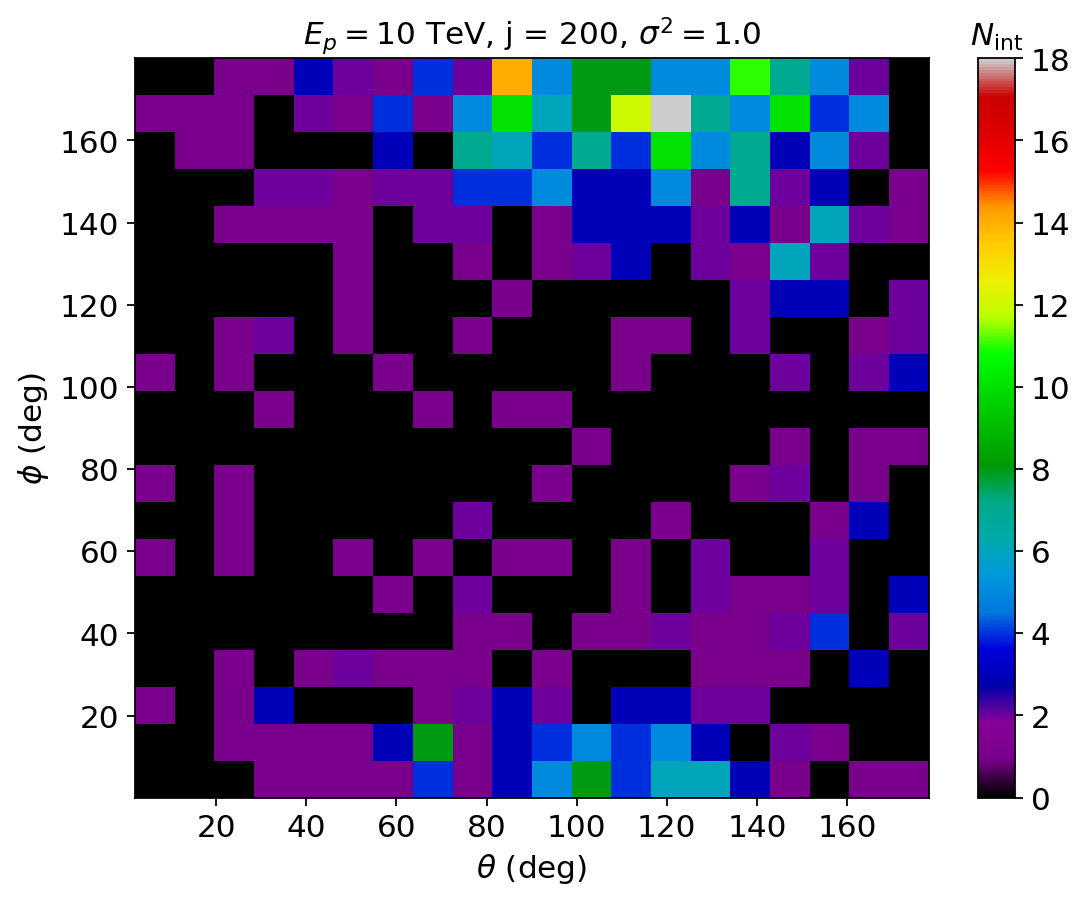} 
    \caption{2D histograms showing the $\theta$ and $\phi$ angles defining the velocity direction of interacting protons injected at $200 < j < 210$, obtained with $\sigma^2 = 0.001$ (left panels) and $\sigma^2 = 1.0$ (right panels), and different energies: $E_p = 100$ GeV in the upper panels, and $E_p = 10$ TeV in the lower panels.}
    \label{fig:angles}
\end{figure*}

\subsection{Angular distribution of outgoing GCRs}
\label{sec:angular}
Gamma-rays produced in the photosphere by GCR proton - ambient proton collisions travel the same direction of the highly relativistic GCR proton, due to momentum conservation, out to Earth at 1 AU (scattering of gamma-ray photons with chromospheric/coronal ambient plasma is neglected here).
The 2D histograms presented in Figure \ref{fig:angles} portray the definition of $\theta$ and $\phi$ angles of the GCR velocity vector (defined in Figure \ref{fig:arcade_3D}), that indicate the directions of the protons at the time of interaction with ambient protons for both the $100$ GeV (upper panels) and $10$ TeV (lower panels) cases. 
The histograms are obtained for $\sigma^2 = 0.001$ (left panels) and for $\sigma^2 = 1$ 
(right panels), with protons injected at the same altitude, specifically $200 < j < 210$.

In the lower energy case (upper panels of Figure \ref{fig:angles}), turbulence plays a crucial role in dictating the direction of the gamma-ray emission.
For weak turbulence (upper left panel of Figure \ref{fig:angles}), the near-isotropic $\theta$-distribution ranges from approximately $0^{\circ}$ to $180^{\circ}$.
Gamma-rays are primarily emitted close to direction grazing the Sun surface (within $0^{\circ}\lesssim \phi \lesssim 60^{\circ}$ or $140^{\circ} \lesssim \phi \lesssim 180^{\circ}$, even if the emission is non-negligible at intermediate $\phi$ values), with a notable concentration in a few red spots on the right (i.e., along semi-axis $x>0$) side (see Figure \ref{fig:int_y_500.png}) of the domain ($\phi \sim 10^{\circ}$) and in proximity to the $x-y$ plane, i.e., plane of the arcade ($80^{\circ}\lesssim \theta \lesssim 140^{\circ}$).
The depletion cone for $\sigma^2 =1$ (upper right panel) shows that emission is preferentially close to the tangent to the Sun surface ($40^{\circ}\lesssim \theta \lesssim 160^{\circ}$), and protons predominantly interact on the left side of the domain ($150^{\circ}\lesssim \phi \lesssim 180^{\circ}$).

Examining the $10$ TeV case (lower panels of Figure \ref{fig:angles}), the effect of turbulence is less evident because the two histograms show a morphology close to each other compared with the two upper plots for the $100$ GeV case. 
Regardless of the value of $\sigma^2$, protons mainly interact within the range $80^{\circ} \lesssim \theta \lesssim 160^{\circ}$, and $140^{\circ} \lesssim \phi \lesssim 180^{\circ}$ or $0^{\circ} \lesssim \phi \lesssim 40^{\circ}$.
This indicates that photons are emitted in proximity to the $x-y$ plane, i.e., plane of the arcade, and, once again, close to the grazing direction. 
In the strong turbulence case (lower right panel of Figure \ref{fig:angles}) the direction of the outgoing protons is skewed toward the left side of the domain ($140^{\circ} \lesssim \phi \lesssim 180^{\circ}$), albeit still in proximity to the $x-y$ plane ($\theta$ not far from $90^\circ$).
In summary, at higher energy and for any turbulence strength, a wide depletion cone around the vertical direction (into the Sun surface) leads to gamma-ray emission close to the plane tangent to the Sun surface, i.e., predominant near the solar limb $(within \ \Delta \phi \sim 40^\circ)$.

In short, weak turbulence narrows the depletion cone at low energies ($E_p=100$ GeV), allowing more isotropic disk emission. At high energy ($E_p=10$ TeV), the turbulence has no effect due to the large $r_g$.


Previous models \citep{Li2024} calculated gamma-ray emission angles for energies ranging from $1$ GeV up to $10^3$ GeV, identifying constant emission at $50^{\circ} \lesssim \theta \lesssim 80^{\circ}$, with $\theta = 90^\circ$ being the direction vertical to the Sun surface.
Our findings indicate that emission angles are energy-dependent, with a broader range of $\theta$ at lower energy. 
This discrepancy may arise from the fact that \cite{Li2024} assumed a different magnetic field geometry.
It seems that achieving a more isotropic emission would necessitate incorporating our component derived from closed magnetic field lines.

\subsection{Disk Gamma-ray flux}
\label{sec:flux}
\begin{figure*}
\centering
\includegraphics[width=0.99\textwidth]{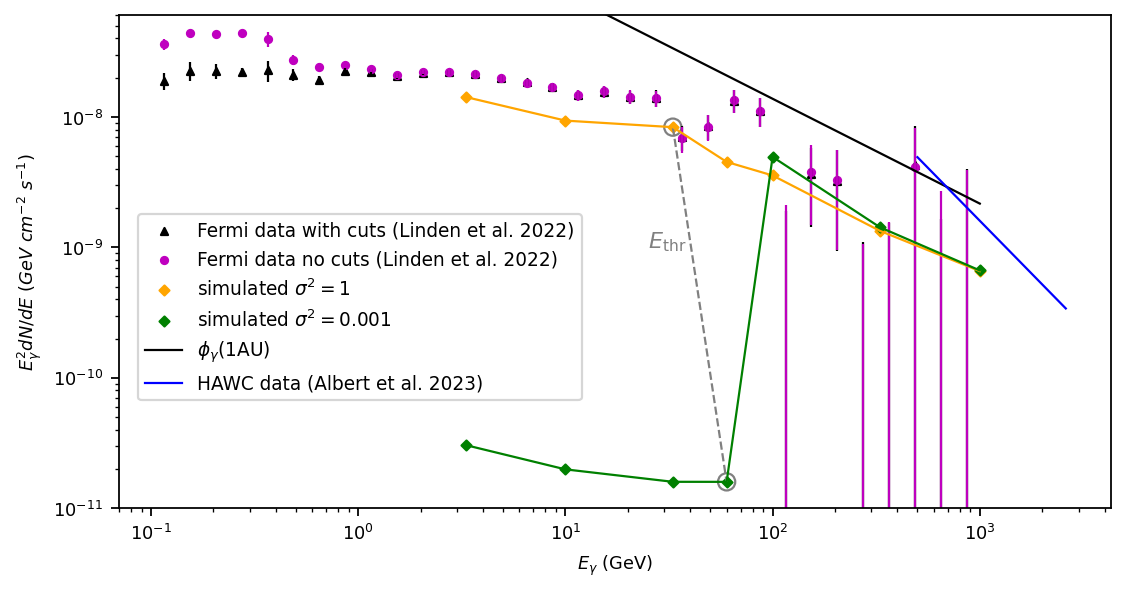 }
\caption{Average solar disk gamma-ray spectrum collected by Fermi-LAT from August 2008 to February 2020 \citep[including the solar flares in the purple points and excluding them in the black points, from][]{Linden2022}, by HAWC \citep[blue line, from][]{Albert2023}, and obtained from our simulations (with $\sigma^2 = 1$ in orange and $\sigma^2 = 0.001$ in green). The dashed gray line marks the threshold energy ($E_\mathrm{thr}$) above which cross-field diffusion drives the GCRs migration into the arcade, depending on $\sigma^2$. The black solid line represents the estimated gamma-ray flux at $1$ AU in the case of maximal efficiency of GCR/gamma-ray conversion (see text for details).}
\label{fig:flux}
\end{figure*}

To validate our model, we calculate the flux of outgoing gamma-rays, comparing the results with observations from both Fermi-LAT and HAWC.
Figure \ref{fig:flux} presents a comparison between the gamma-ray flux detected by Fermi-LAT \citep[purple and black points, from][]{Linden2022} and HAWC \citep[blue solid line, from][]{Albert2023}, and the flux obtained from our simulations (orange and green points).  
The two Fermi-LAT datasets differ in that the black dataset excludes flares, while the purple one represents the complete dataset without cuts.
The green points correspond to the numerically calculated solar disk flux (Eq. \ref{eq:flux:final}) in the weak turbulence case ($\sigma^2 = 0.001$), while the orange points correspond to strong turbulence ($\sigma^2 = 1$). Each set of points in Figure \ref{fig:flux} assumes a fixed value of $\sigma^2$ over the solar disk, although the observed flux is certainly contributed by structures that span a range of turbulence strength, not considered here for the sake of simplicity.
We apply the correction $\xi$ for the gamma-ray flux due to the energy dependence of the cross-field diffusion from open to closed field lines at low energies and only up to energies ($E_\mathrm{thr}$, see gray dashed line) where cross-field diffusion is likely to suppress GCRs migration into the arcade, hence gamma-ray flux (refer to Section \ref{sec:diff_coeff} for further details).
The black solid line represents the estimated gamma-ray flux at $1$ AU, using the expression $\phi_\gamma (1AU) \approx (E_p^2 dN/dE_p)_\mathrm{obs} (R_\odot/R_\mathrm{1AU})^2(E_\gamma/E_p)$, where $E_\gamma/E_p \sim 1/10$ as mentioned previously.



Figure \ref{fig:flux} shows that our numerically computed flux for $\sigma^2 = 1$ well explains both the slope and absolute value with the Fermi/LAT observed spectrum, whereas the contribution to the flux from weaker turbulent structures is suppressed at energies $E_\gamma < 100$ GeV by the cross-field diffusion. We note that the contribution to the gamma-ray flux of more laminar arcades ($\sigma^2 \ll 1$), despite negligible at energies $E_\gamma \lesssim 100$ GeV (see Fig. \ref{fig:flux}), has to be folded in the calculation of the total disk flux as it can play a role in the spectral dip (see below). 

The weak turbulence gamma-ray flux (represented by green points) exhibits a discontinuity at the energy threshold $E_\mathrm{thr}$ (see Eq. \ref{eq:deltax}) determined by $\sigma^2$.  
If the gamma-ray spectrum is calculated by incrementing the turbulence strength $\sigma^2$ from $0.001$ up to $1.0$, this discontinuity progressively shifts to lower energies due to the dependence of the energy threshold on $\sigma^2$ (see Eq. \ref{eq:deltax}). 
Although our study assumes uniform turbulence strength across all solar loop structures, by folding the contributions of structures characterized by distinct $\sigma^2$ values, hence leading to discontinuities in the gamma-ray flux at different energy thresholds between about $33$ and $100$ GeV, we speculate that the dip/rebrightening of the LAT gamma-ray flux can originate from a combination of structures with a certain range of $\sigma^2$ spread over the solar disk. Thus, rather than a dip, the combined effect of local turbulence would originate a bump in the high-energy tail of the flux.
At higher energies, the presence of turbulence does not truly influence the gamma-ray flux in Figure \ref{fig:flux}.
Indeed, at $E_\gamma = 10^3$ GeV, the gamma-ray flux obtained with the two different values of $\sigma^2$ overlap, consistent with the trend seen in the right panel of Figure \ref{fig:trend_int}, where the number of interacting particles remains relatively constant with $\sigma^2$.
This underscores the indispensable role of turbulence in accurately modeling the observed gamma-ray spectrum.


Our computed gamma-ray flux exhibits a steeper slope compared to the incoming GCR proton spectrum. 
Specifically, for photon energies exceeding $33$ GeV, we obtain an approximate slope of $-0.8$, whereas the cosmic ray proton spectrum follows a decline of approximately $E_p^{-0.7}$ \citep[see, e.g.,][]{Blasi2013, Workman2022}.
This result underscores the discernible impact of magnetic arcades on GCRs motion.

Overall, the slope of the solid black line in Figure \ref{fig:flux}, i.e., gamma-ray flux at $1$ AU, corresponding to the maximal efficiency case of one photon emitted per incoming GCR, is comparable to the one of the Fermi-LAT data at high energies ($E_\gamma \gtrsim 100$ GeV); however, a notable deviation is observed at lower energies. 
This flattening suggests that trapping by the magnetic arcade might contribute significantly to the shaping of the gamma-ray emission at lower energies, due to the smaller gyroradius.

\section{Discussion and conclusions}
\label{sec:summary}
%
%

This study presents numerical simulations of test-particle protons within a turbulent magnetic field arcade extending from the bottom of the photosphere up to $0.03$ solar radii. 
Utilizing the PLUTO code, computations are conducted across various turbulence strengths.
The primary objective is to assess the impact of these closed magnetic field structures on the gamma-ray emission from the solar disk.

Our analysis indicates that protons primarily interact in the lower section of the arcade structure, characterized by higher density and shorter interaction times, in agreement with \cite{Li2024}.
As turbulence increases ($\sigma^2 = 1$), cross-field diffusion \citep[e.g.,][]{Fraschetti2011}  becomes more significant, causing particles to traverse magnetic field lines and approach the high-density low atmosphere at the center of the arcade. However, at low-energies ($E_p \lesssim 100$ GeV), cross-field diffusion does not suffice to enhance GCR-proton collisions and, together with the motion along the z-axis ($\delta B_z \neq 0$), facilitates counterintuitively the escape from the arcade with a net loss of $N_{int}$ as $\sigma^2$ increases.
In cases of high-energy particles ($E_p \gtrsim 10$ TeV), where the gyroradius is larger, turbulence plays a negligible role in determining interaction and escape rates.
Despite the reduction of $N_\mathrm{int}$ with $\sigma^2$, our findings emphasize the primary role of turbulence in allowing access of GCRs into the magnetic arcade structure from adjacent open flux tubes.

Our investigation reveals an energy-dependent directional pattern in gamma-ray emission from the solar disk, modulated by turbulence strength. 
At lower energies, the emission at the disk center remains non-negligible. 
In contrast, at higher energies, turbulence exerts minimal influence, and the emission is concentrated near the solar limb.

Turbulence ($\sigma^2 = 1$) plays a crucial role in accurately modeling the observed gamma-ray flux, albeit resulting in a calculated intensity slightly lower by a factor of $\lesssim 2$, as well as the difference between the gamma-ray flux at solar minimum and maximum \citep{Linden2022}.
The gamma-rays emitted within the flux tubes, as investigated by \cite{Li2024}, add to the intensity of the low-energy ($\lesssim 10$ GeV) flux.
Addressing the energy-dependent cross-field transport of GCRs from open to closed field lines flattens the calculated gamma-ray flux at lower energies, consistent with Fermi-LAT observations \citep[as detailed in][]{Linden2022}.
However, at lower energy and weaker turbulence ($\sigma^2 = 0.001$), the gamma-ray flux is far lower, as the penetration into the arcade of most GCRs is inhibited.

Our model overcomes the conclusions drawn in SSG91, in that we predict a higher flux consistently with Fermi-LAT observations, and it also extends to the higher energies observed by HAWC \citep{Albert2023}.
By convolving the gamma-ray spectrum obtained from the magnetic arcades over the entire solar disk, with a certain distribution of turbulence strengths $\sigma^2$, we speculate that the re-brightening (also known as spectral dip, see the last paragraph of Sect. \ref{sec:puzzles}) in the gamma-ray flux observed by Fermi-LAT within the energy range of approximately $33-100$ GeV can be explained.

The hadronic origin of gamma-rays from the solar disk is expected to largely dominate over proton synchrotron emission.
The power emitted by a single proton is concentrated around the characteristic synchrotron energy $\epsilon_s$, as expressed by $\epsilon_s = 0.29 (3 e h \gamma^2 B)/(4 \pi m_p c)$ \citep{Longair:11,Fraschetti2018}.
Here, $e$ represents the electric charge, $h$ is the Planck constant, and $\gamma$ is the GCR proton Lorentz $\gamma$-factor. For instance, within a magnetic field strength of $B = 100$ Gauss, a proton is capable of emitting a photon with energy $\epsilon_s \gtrsim 1$ GeV (consistent with the Fermi-LAT regime) through the synchrotron mechanism only at  energies $E_p = m_p c^2 \gamma \gtrsim 1.8 \times 10^6$ TeV.
At these high energies, the GCR proton flux observed at $1$ AU is far lower than the GCR protons flux \citep{Margiotta.etal:14} needed to produce $E_\gamma = 1$ GeV, i.e., $E_p \sim 10\, E_\gamma = 10$ GeV (see Sect. \ref{sec:cross_section}).
In addition, within the solar disk  the hadronic-origin gamma-rays dominate over the IC emission (see Appendix \ref{app:emission} for further details).
However, as argued by \cite{Orlando2023}, synchrotron emission by GCR electrons could be used to probe the magnetic field of the solar atmosphere.

Throughout this study, we assumed a pure protons composition for both the incoming GCRs flux and the solar atmosphere, neglecting heavier ions species.
Indeed, the Helium abundance, $\lesssim 10 \%$ in the GCR \citep[see, e.g.,][]{Gaisser2016} and in the solar atmosphere \citep[see, e.g.,][]{Asplund2009, Moses2020}, is expected to increase the gamma-ray flux by a factor $< 2$ \citep{Zhou2017,Li2024}. 
\cite{Rankin2022} analyzed Parker Solar Probe (PSP) in-situ measurements and observed a radial intensity gradient between $0.1$ and $1$ AU for Anomalous Cosmic Rays heavy ions $\simeq 49.4\%$/AU. The PSP-measured radial gradient is much larger than model predictions \citep[see, e.g.][]{Strauss2010} below $2$ AU.
Although PSP measurements probe much smaller CR energies (i.e., $6.9–27$ MeV nuc$^{-1}$), it is not unreasonable to extrapolate that current innermost heliosphere data suggest that the abundance of heavy ions ($> 10$ GeV/nuc) GCR impinging on the Sun might be much lower than expected, thus not contributing significantly to the Fermi/LAT gamma-ray flux. 
Since the uncertainty on the filling factor of the arcade is likely to be larger than $2$ and time-dependent across the solar cycle, and we focus here on the solar cycle-averaged $\gamma$-spectral shape only, a multi-species composition is unlikely to modify the calculated spectrum.

Throughout the course of this investigation, modifications have been made to both the magnetic field and solar atmospheric models (ambient density).
Variations in magnetic field strength $B_0$ and magnetic scale height $\Lambda_B$ surprisingly did not alter the slope of the resulting gamma-ray spectrum but rather affected its overall intensity.
Specifically, an increase in $B_0$ and $\Lambda_B$ correlated with increased gamma-ray flux.
Due to the lack of solid observational constraints on the vertical profile of the photospheric plasma density, to our knowledge, we have implemented a number of different profiles. 
An exponentially decreasing density profile ($\rho = \rho_0 e^{-y/\Lambda_B}$) across the photosphere, mimicking the exponential decrease of the chromospheric magnetic field, leads to discrepancies between the
observed flux and the modeled one, with this density profile also conflicting with solar atmospheric modeling.
Numerically calculated density profile via MHD resistive simulations agree on the general shape, although differ significantly in the normalization at the base of the photosphere. We adopted here the 
density profile from \cite{Morton2023} for the low photosphere and up to 4 Mm, and from \cite{Gonzales2021} at higher altitudes (up to $30$ Mm). The latter density profile is consistent with 
the possibility of supra-thermal electrons exciting the maser instability at the top of the arcade  \citep[see][]{Winglee1986}.
The computational domain is extended to match the bottom of the Sun's photosphere due to the density difference between the bottom and top of the photosphere, which drastically influences the interaction times and causes a disagreement in gamma-ray flux with observations.
Expanding the computational box by a factor of ten with proportionally increased scale height led to an increase of the trapping effect, especially for high-energy particles with larger gyroradii. This resulted in increased particle interactions and a higher flux.
However, such large magnetic structures ($\sim 2 \times 10^5$ km) are either unlikely to be stable or very rare, as evidenced by recent Solar Orbiter observations \citep{Antolin.etal:23}.

Significantly, the computed gamma-ray flux deviates from the estimated gamma-ray flux at $1$ AU, calculated under the assumption of full efficiency in converting GCR protons into gamma-rays.
This discrepancy underscores the impact of magnetic arcades on the gamma-ray flux.
This inference finds support in the result that the crossing time of the arcade in the vertical direction consistently falls below the peak of the interaction time distribution (not shown here). 
It follows that magnetic arcades possess the capability to temporarily confine particles, thereby exerting influence over the interaction mechanism and subsequently modulating gamma-ray emissions. 
In future investigations, our focus will shift towards evaluating the trajectory of GCR protons as they transition from the flux tubes (i.e., open magnetic field lines) to these closed magnetic field loops.

\section*{}
\noindent
The authors express gratitude for the valuable comments provided by the referee, as well as the insightful feedback from Dr.s Asgari-Targhi, J. Beacom, J.-T. Li, T. Linden, A. Peter, X. Zhao. E.P. was supported by NASA under grant 80NSSC22K0040. F.F. was partially supported by NASA under grants 80NSSC22K0040, 80NSSC18K1213, 80NSSC21K0119 and 80NSSC21K1766. F.F. was also supported, in part, by NASA through Chandra Theory Award No. TM0-21001X, TM6-17001A issued by the Chandra X-ray Observatory Center, which is operated by the Smithsonian Astrophysical Observatory for and on behalf of NASA under contract NAS8-03060. 
J. K. and J. G. were partially supported by NASA under grant 80NSSC22K0040.

%

\vspace{5mm}
\facilities{AMS, CREAM, Fermi-LAT, HAWC}


\software{PLUTO code \citep{Mignone2007, Mignone2012}
          }



\appendix

\section{Leptonic emission}
\label{app:emission}

We compare below an alternative leptonic process (IC scattering) for the gamma-rays production in the solar atmosphere with the hadronic process calculated above.

Solar UV photons are scattered off high-energy electrons (or positrons) via IC and boosted into the gamma-ray band. 
Below we consider the Klein-Nishina regime, holding if $4 \epsilon_0 \gamma/m_ec^2 \gg 1$, where $\epsilon_0 = 5$ eV is the 
UV initial photon energy \citep[see, e.g.,][]{Fraschetti2023}.


The IC power emitted by a single electron is equal to
\begin{equation}
\label{eq:hadronic}
    P_\mathrm{IC}^{KN} \simeq \frac{3}{8} \sigma_T c (m_e c^2)^2 \frac{n_0}{\epsilon_0}\left[\ln \left(\frac{4 \epsilon_0 \gamma}{m_ec^2}\right) - \frac{11}{6}\right],
\end{equation}
where $\sigma_T$ is the Thomson cross-section and $n_0 = \sigma T^4/\epsilon_0 c$ is the ambient photon number density \citep[see, e.g.,][]{Schlickeiser:09,Fraschetti2017}, with $\sigma$ being the Stefan-Boltzmann constant and $T \sim 5000$ K the photosphere temperature. We have assumed that in the low solar photosphere, where the IC might occur, the photon distribution is nearly-isotropic, in contrast with the dominantly outgoing photons in the low corona.

\begin{figure}
\centering
\includegraphics[width=0.55\textwidth]{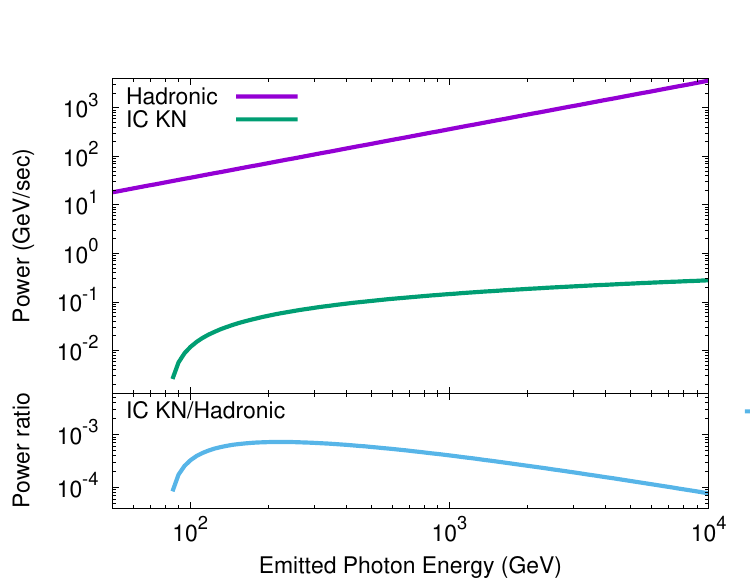}
\caption{\textit{Top panel}: Power emitted by a single electron in the Klein-Nishina regime (green line) compared with the power emitted by hadronic proton-proton collision (purple line) investigated in detail in the present work. \textit{Bottom panel}: Ratio of the single electron IC power emitted in the Klein-Nishina regime to the power emitted by proton-proton collision (light blue line).}
\label{fig:IC}
\end{figure}

Using Equation 2.47 ($E_\gamma \simeq E_e$) from \cite{Blumenthal1970} for the Klein-Nishina regime, the electron's energy can be converted into photon energy. The upper panel of Figure \ref{fig:IC} illustrates the power emitted in the Klein-Nishina (green line) regime as a function of the emitted photon energy. 
The purple line in this panel represents the power emitted by a single proton through hadronic 
collisions modeled in detail in this paper, that can be simplified as
\begin{equation}
    P_\mathrm{pp} \simeq c \sigma_\mathrm{pp}n_p \bar{F_\gamma}E_\gamma,
\end{equation}
where $\sigma_\mathrm{pp} \sim 30$ mb is the inelastic cross-section of the proton-proton interaction, $n_p \sim 10^{15}$ cm$^{-3}$ denotes the lower limit of the 
plasma density within the arcade necessary for the gamma-ray emission, 
while $\bar{F_\gamma} \simeq 4$ is the value at the peak for $E_p = 1$ TeV and $E_\gamma/E_p \simeq 0.1$ \citep[see Figure 6 from][]{Kelner2006}.
The lower panel of Figure \ref{fig:IC} illustrates the ratio between the Klein-Nishina power and the hadronic  power in the maximal efficiency case, i.e., every incoming GCR produces a gamma-ray (in light blue).
In addition, in the dense low photosphere, $n_p$ 
is considerably higher ($n_p \sim 10^{18}$ cm$^{-3}$, see Sect. \ref{sec:box_size}), and the power ratio drops below $10^{-4}$. 
These results confirm that, as expected, hadronic emission from closed arcades permeating the solar surface dominates the gamma-rays emission from the solar disk. 

\bibliography{sample631}{}
\bibliographystyle{aasjournal}



\end{document}